\documentclass{aa}

\usepackage{amssymb}
\usepackage{natbib}
\usepackage{mathtools}
\usepackage[version=4]{mhchem}
\usepackage{siunitx}
\usepackage{multirow}
\usepackage{lipsum}
\usepackage{blindtext}
\usepackage{array}
\usepackage{subcaption}
\usepackage{fixltx2e}
\usepackage{gensymb}
\usepackage{placeins}
\newcolumntype{H}{>{\setbox0=\hbox\bgroup}c<{\egroup}@{}}

\begin{document}
\title{Molecular shock tracers in NGC 1068: SiO and HNCO}

\author{G. Kelly \inst{\ref{inst1}}\and S. Viti\inst{\ref{inst1}}\and  S. Garc\'ia-Burillo \inst{\ref{inst2}}\and A. Fuente\inst{\ref{inst3}}\and  A. Usero \inst{\ref{inst2}} \and M. Krips \inst{\ref{inst5}}\and  R. Neri \inst{\ref{inst5}} }
\institute{Department of Physics and Astronomy, University College London, Gower Street, London, WC1E 6BT, U.K. email{: g.kelly@ucl.ac.uk}\label{inst1} 
\and 
Observatorio Astron\'omico Nacional (OAN)-Observatorio de Madrid, Alfonso XII 3, 28014-Madrid, Spain\label{inst2}
\and
Observatorio Astron\'omico Nacional (OAN, IGN), Apdo 112, 28803 Alcal\'a de Henares, Spain\label{inst3}
\and
Institut de Radioastronomie Millim\'atrique, Domaine Universitaire, 38406 St-Martin-d'H\`eres, France\label{inst5}}

\abstract{}{We present and compare the distribution of two shock tracers, SiO and HNCO, in the Circumnuclear Disk (CND) of NGC 1068. We aim to determine the causes of the variation in emission across the CND.}{SiO$(3-2)$ and HNCO$(6-5)$ emission has been imaged in NGC 1068 with the Plateau de Bure Interferometer (PdBI). We perform an LTE and RADEX analysis to determine the column densities and physical characteristics of the gas emitting these two lines. We then use a chemical model to determine the origin of the emission.}{There is a strong SiO peak to the East of the AGN, with weak detections to the West. This distribution contrasts that of HNCO, which is detected more strongly to the West. The SiO emission peak in the East is similar to the peak of the molecular gas mass traced by CO. HNCO emission is offset from this peak by as much as $\sim$80 pc ($\leqslant$1$''$). We compare velocity integrated line ratios in the East and West. We confirm that SiO emission strongly dominates in the East, while the reverse is true in the West. We use RADEX to analyse the possible gas conditions that could produce such emission. We find that, in both East and West, we cannot constrain a single temperature for the gas. We run a grid of chemical models of potential shock processes in the CND and find that SiO is significantly enhanced during a fast (60 km s$^{-1}$) shock but not during a slow (20 km s$^{-1}$) shock, nor in a gas not subjected to shocks at all. We find the inverse for HNCO, whose abundance increases during slow shocks and in warm non-shocked gas.}{High SiO and low HNCO indicated a fast shock, while high HNCO and low SiO indicates either a slow shock or warm, dense, non-shocked gas. The East Knot is therefore likely to contain gas that is heavily shocked. From chemical modelling, gas in the West Knot may be non-shocked, or maybe undergoing a much milder shock event. When taking into account RADEX results, the milder shock event is the more likely of the two scenarios.}

\maketitle

\section{Introduction}

Molecular emission is now routinely used to probe and trace the physical and chemical processes in external galaxies.
Over the past 20 years or so, different molecules have been found to trace different gas components within a galaxy (for example HCO, HOC$^+$ in PDRs (e.g. \citealt{2004ApJ...616..966S, 2002ApJ...575L..55G}), HCN and CS for dense gas (e.g. \citealp{2004ApJ...606..271G, 2008ApJ...685L..35B})). However, it is seldom possible to identify one molecular species with one gas component only, as often energetics play a key role in shaping the spectral energy distribution of the molecular ladders. Of particular interest to this study are the molecules SiO and HNCO, which are both well known tracers of shocks \citep{1997ApJ...482L..45M, 2010A&A...516A..98R}. Both these molecules have been observed and used as shock tracers in external galaxies \citep{2005ApJ...618..259M, 2006A&A...448..457U, 2009ApJ...694..610M, 2010A&A...519A...2G}. 

HNCO may be formed mainly on dust grain mantles \citep{2015MNRAS.446..439F} or possibly in the gas phase, followed by freeze out to the icy mantles \citep{2015MNRAS.449.2438L}. In either cases its location on the outer regions of the dust grain means that it is easily sublimated even in weakly shocked regions; hence HNCO may be a particularly good tracer of low velocity shocks. Silicon, on the other hand, is partially depleted from the gas to make up the dust grain itself. This extra silicon is only released into the gas phase in higher velocity shocked regions through sputtering. Once it is in the gas phase, it can react with molecular oxygen or a hydroxyl radical to form SiO \citep{1997A&A...321..293S}, which can then be used to trace more heavily shocked regions. Thus HNCO and SiO concomitant detection in a galaxy where shocks are believed to take place may be able to give us a fuller picture of the shock history of the gas.

NGC 1068 is a well-studied nearby (D = 14 Mpc \citep{1997Ap&SS.248....9B}, 1$''$ $\approx$ 70 pc) Seyfert 2 galaxy. Its molecular gas is distributed over three regions \citep{2000ApJ...533..850S}: a starburst ring with a radius $\sim$1.5 kpc, a $\sim$2 kpc stellar bar running north East from a circumnuclear disk (CND) of radius $\sim$200 pc. {\citet{2010A&A...519A...2G} used the Plateau de Bure Interferometer (PdBI) to map the galaxy and found strong detections of SiO$(2-1)$ in the East and West of the CND. The SiO kinematics of the CND point to an overall rotating structure, distorted by non-circular and/or non-coplanar motions. The authors concluded that this could be due to large scale shocks through cloud-cloud collisions, or through a jet-ISM interaction. However, due to strong detections of CN not easily explained by shocks, they also suggest that the CND could be one large X-ray dominated region (XDR)}. More recently, the CND has been mapped at very high resolution with ALMA in several molecular transitions \citep{2014A&A...567A.125G, 2015ASPC..499..109N, 2016ApJ...823L..12G}. In Garcia-Burillo et al. (2014), five chemically distinct regions were found to be present within the CND: the AGN, the East Knot, West Knot and regions to the north and south of the AGN (CND-N and CND-S). Viti et al. (2014) combined these ALMA data with PdBI data and determined the physical and chemical properties of each region using a combination of CO rotation diagrams, LVG models and chemical modelling.  It was found that a pronounced chemical differentiation is present across the CND and that each sub-region could be characterised by a three-phase component interstellar medium: one of these components comprises shocked gas and seems to be traced by { a high--J (7--6) CS} line. 

We now resolve the CND in both SiO$(3-2)$ and HNCO$(6-5)$ and couple these with previous lower-J observations. In Section 2 we describe the observations, while in Section 3 we present the molecular maps. In Section 4 we present the spectra at each location. In Section 5 we perform an LTE and RADEX analysis in order to constrain the physical conditions of the gas, as well as chemical modelling to determine its origin. We briefly summarise our findings in Section 6.

\section{Observations}\label{PdbI}

Observations of NGC\,1068 were carried out with the PdBI array \citep{1992A&A...262..624G} between 2010 
January 23 and March 5. We used the AB configurations and six antennae. We simultaneously observed the 
($v$=0, $J$=3--2) line of SiO (at 130.2686\,GHz) and the $J$=6--5 band of HNCO (at 131.4--132.4\,GHz). The 
HNCO(6--5) band is split up into many hyperfine lines blended around five groups ($(K_{\rm p},K_{\rm o})$=$(1,6)-(1,5), (2,5)-(2,4), (2,4)-(2,3), (0,6)-(0,5)$, and $(1,5)-(1,4)$). 
The strongest group of lines, ($(K_{\rm p},K_{\rm o})$=$(0,6)-(0,5)$), lies at 131.8857\,GHz. During the 
observations the spectral correlator was centered at the mock rest frequency 131.0765\,GHz.  This choice 
allowed us to cover simultaneously the SiO and HNCO lines. Rest frequencies were corrected for the recession 
{ heliocentric} velocity initially assumed to be $v_{o}(HEL)$=1137\,km~s$^{-1}$.  The correlator configuration covers a 
bandwidth of 1\,GHz for this setup, using eight 320\,MHz-wide units with an overlap of 90\,MHz; this is 
equivalent to 2290\,km~s$^{-1}$ at 131.0765\,GHz. { The phase and amplitude calibrators were 0215+015 and 0138-097, with fluxes accurate to 10\%. The flux calibrator was MWC349 giving 1.36 Jy and a stable flux with an accuracy of 5\%.}. Observations were conducted in single pointing mode { with a field-of-view of} 
size 38.8$^{\prime\prime}$  centered at $\alpha_{2000}$=02$^{h}$42$^{m}$40.71$^{s}$ and $\delta_{2000}$=--00$^{\circ}$00$^{\prime}$47.94$\arcsec$. The latter corresponds to the nominal position of the 
AGN core, as determined from different VLA and VLBI radio continuum images of the galaxy (e.g. \citealt{1996ApJ...458..136G}). Visibilities were obtained through on-source integration times of 20 minutes framed by short ($\sim$\,2\,min) phase and amplitude calibrations on nearby quasars. The absolute flux scale in our maps was 
derived to a 10$\%$ accuracy based on the observations of primary calibrators whose fluxes were determined 
from a combined set of measurements obtained at the 30m telescope and the PdBI array. 

The image reconstruction was done with the
standard IRAM/GILDAS software \citep{2000ASPC..217..299G}. We used natural weighting and no taper to 
generate the SiO and HNCO line maps with a size of 133$\arcsec$ and 0.13$\arcsec$/pixel sampling; the 
corresponding synthesized beam is $1.1''\times 0.7''$, $PA$=23$^{\circ}$. The conversion factor between Jy\,beam$^{-1}$ and K is 90\,K~Jy$^{-1}$~beam. The point source sensitivities were derived from emission-free 
channels. They are 1.2\,mJy/beam in 5~MHz-wide channels. Images of the continuum emission of the 
galaxy, not shown in this paper, were obtained by averaging those channels free of line emission at both 
frequency ranges. 
 
We have also used the SiO($v$=0, $J$=2--1) PdBI map of \citet{2010A&A...519A...2G} to derive the 
3--2/2--1 line ratio map in T$_{\rm mb}$ units at the common (lower) spatial resolution of  $3.6''\times 2.1''$, 
$PA$=29$^{\circ}$ of the 2--1 line map. This required the use of a Gaussian kernel of $3.4''\times 1.9''$, $PA$=29$^{\circ}$ to convolve the SiO(3--2) map. Hereafter, velocities are referred to $v_{o}(HEL)$=1137\,km~s$^{-1}$.

\begin{table*}[htbp]
\begin{center}
\caption{Locations of peak emission for each line, with offset from CO$(3-2)$ emission}
\label{tab:positions}
\begin{tabular}{c c c c c c}
Line & Position & RA & Dec & $\Delta$RA, $\Delta$Dec ($''$) & Location \\
\hline
SiO$(3-2)$ & East Knot & 02:42:40.771 & -00:00:47.98 & 0, 0.1 & East Knot 1\\
HNCO$(6-5)$ & East Knot & 02:42:40.763 & -00:00:48.60 & -0.1, -0.8 & East Knot 2 \\
CO$(3-2)$ & East Knot & 02:42:40.771 & -00:00:47.84 & 0, 0 & - \\
\hline
SiO$(3-2)$ & West Knot & 02:42:40.608 & -00:00:48.43 & -0.3, -0.6 & West Knot 1\\
HNCO$(6-5)$ & West Knot & 02:42:40.590 & -00:00:48.77 & -0.6, -0.9 & West Knot 2\\
CO$(3-2)$ & West Knot & 02:42:40.630 & -00:00:47.84 & 0, 0 & -\\
\end{tabular}
\end{center}
\end{table*}

\section{Molecular gas maps}

\subsection{SiO}

\begin{figure*}[bhtp]
\centering
{\includegraphics[trim={1cm 6cm 0cm 0cm},clip, width=0.6\linewidth, angle=90]{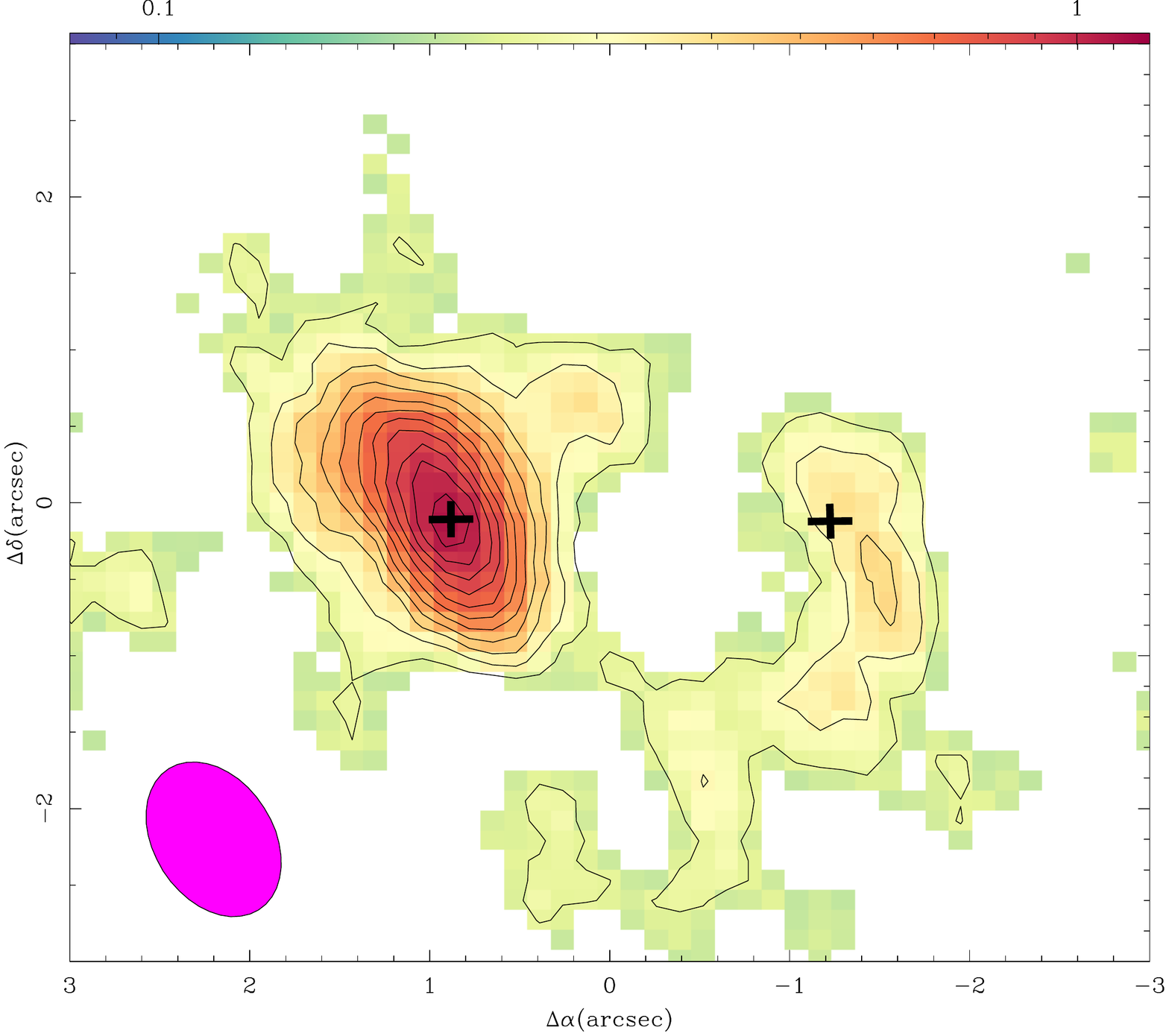}}
\caption{SiO$(3-2)$ map { over the central 6$'' \times$ 6$''$ region}. Contours are from 3$\sigma$ in increments of 1$\sigma$ where $\sigma$ $\sim$0.08 Jykm/s/beam. {The black crosses indicate the approximate position of the CO$(3-2)$ peak. The beam size is displayed in the bottom left.}}
\label{fig:siomap}
\end{figure*}

Figure ~\ref{fig:siomap} shows the SiO$(3-2)$ intensity map over the central 6$'' \times$ 6$''$ region. Data are included across a 500 km s$^{-1}$ spectral window centred around the SiO transition. We see strong emission to the East of the AGN, with weaker detections to the south and West. In agreement with SiO$(2-1)$ \citep{2010A&A...519A...2G}, little emission is seen around the AGN itself. Table ~\ref{tab:positions} shows the coordinates of the peak emission of SiO$(3-2)$ to both the East and West of the AGN. This is compared to those in \citet{2014A&A...570A..28V} from CO$(3-2)$, which is used to identify an East Knot and a West Knot. This notation is continued in this paper.
 
We see that the peak SiO$(3-2)$ emission matches well with the molecular gas traced by CO$(3-2)$ in the East Knot (see Table~\ref{tab:positions}) but it is slightly offset in the West Knot. {It should be noted that since emission here is weak and extended and the beam size is 1$''$, this offset may not be meaningful.} We do not see any noticeable emission outside of the circum-nuclear region about the AGN. The starburst ring is not detected { in SiO$(3-2)$}. 

\begin{figure}[bhtp]
{\includegraphics[width=1.2\linewidth, angle=0]{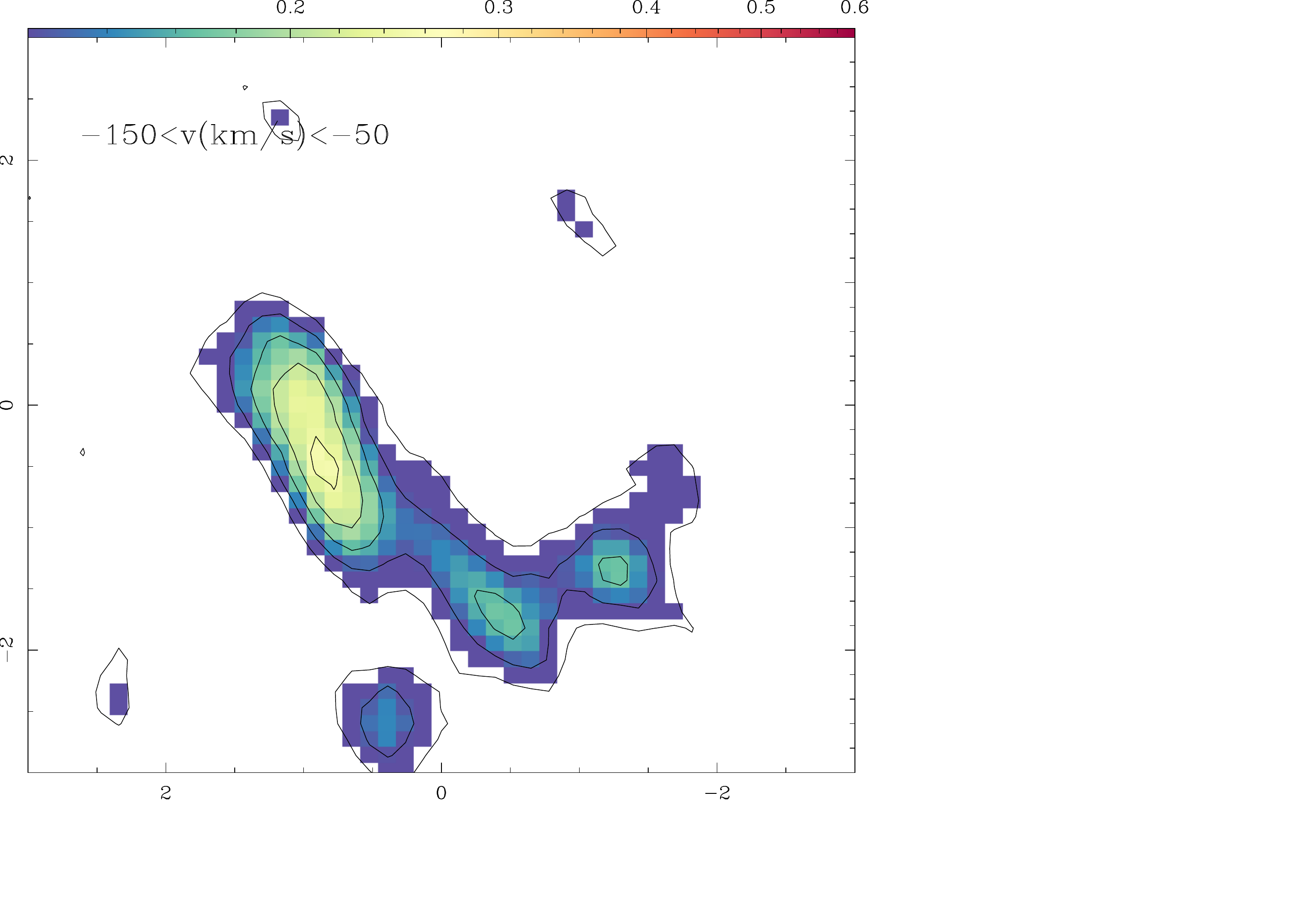}}
{\includegraphics[width=1.2\linewidth, angle=0]{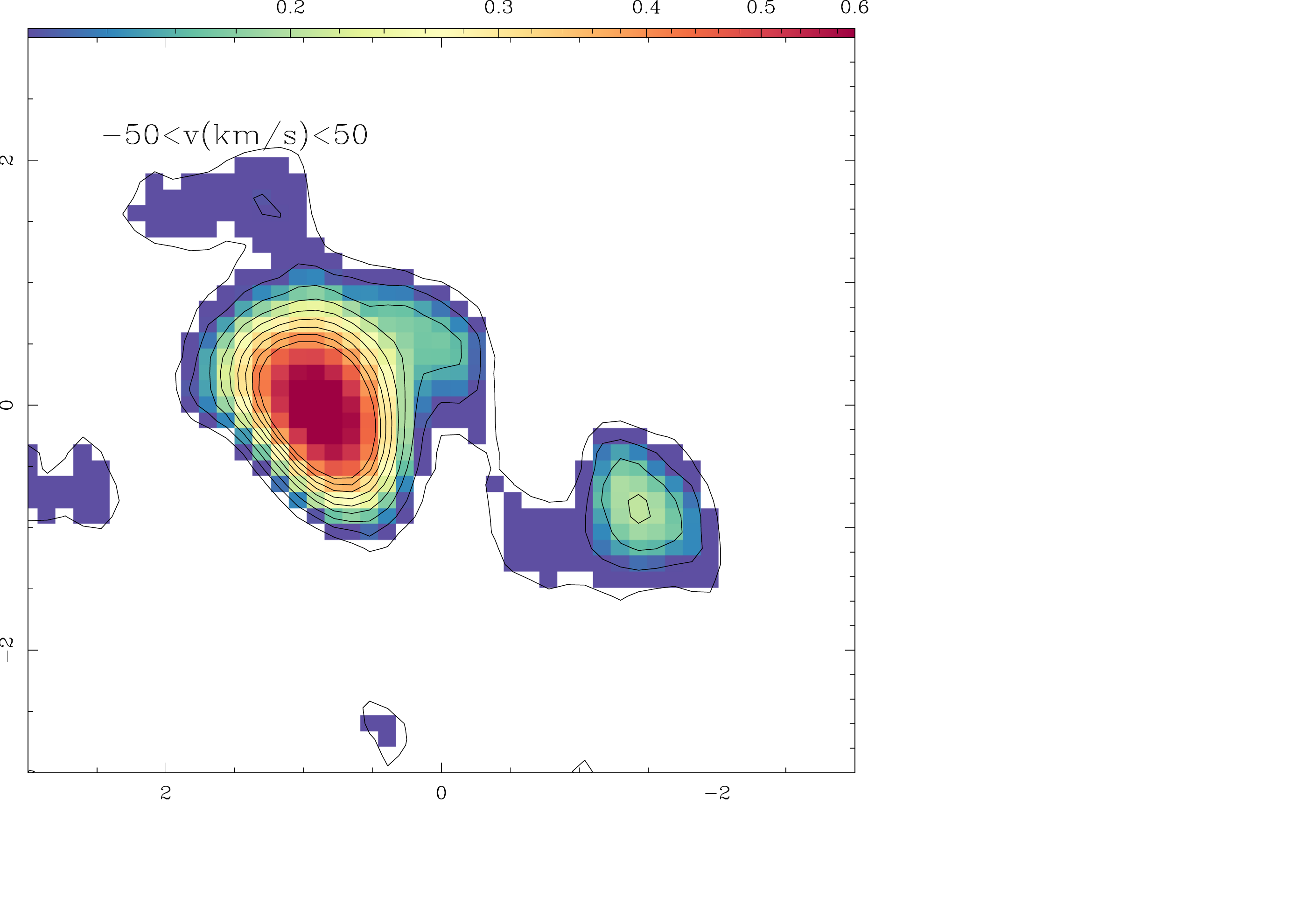}}
{\includegraphics[width=1.2\linewidth, angle=0]{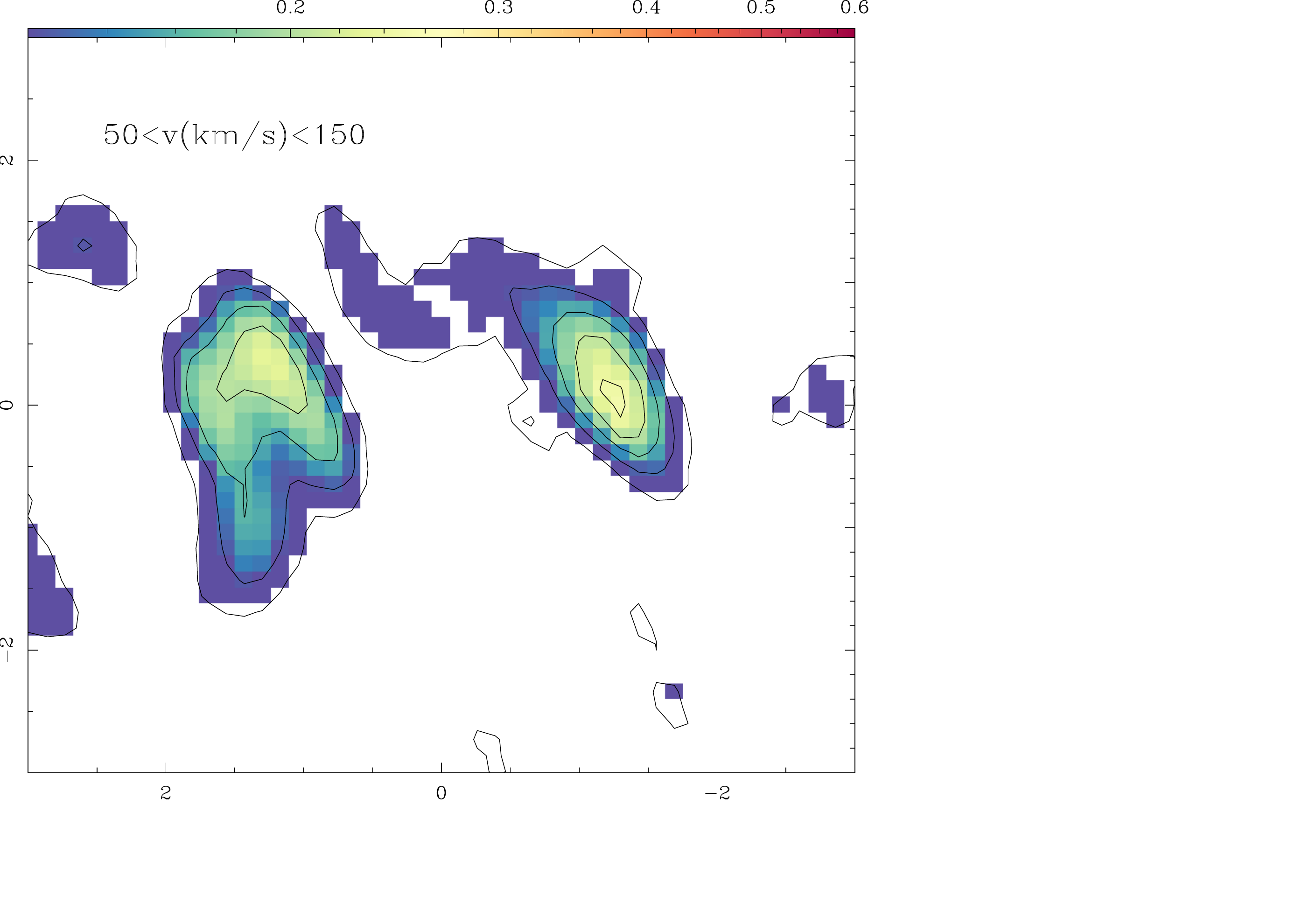}}
\caption{{SiO$(3-2)$ maps for velocity widths of 100 km/s. Contours are from 3$\sigma$ in increments of 1$\sigma$  where $\sigma$ $\sim$0.04 Jykm/s/beam}}
\label{fig:siovel}
\end{figure}

We also present SiO maps for { different velocity bandswidths of 100 km s$^{-1}$, centred around v - $v_{o}$ = 0 km s$^{-1}$  (Figure ~\ref{fig:siovel})}. Examining these plots, it is clear that the majority of SiO emission in the East Knot is coming from the central 100 km s$^{-1}$, with minor contributions from \textgreater 50 km s$^{-1}$ and \textless -50 km s$^{-1}$. { However, in the West Knot, the majority of SiO emission is from velocities  \textgreater 50 km s$^{-1}$. It should be noted, however, that emission here is quite weak}.  This is further analysed in section ~\ref{spec}.

\subsection{HNCO}

\begin{figure*}[bhtp]
\centering
{\includegraphics[trim={1cm 6cm 0cm 0cm},clip, width=0.6\linewidth, angle=90]{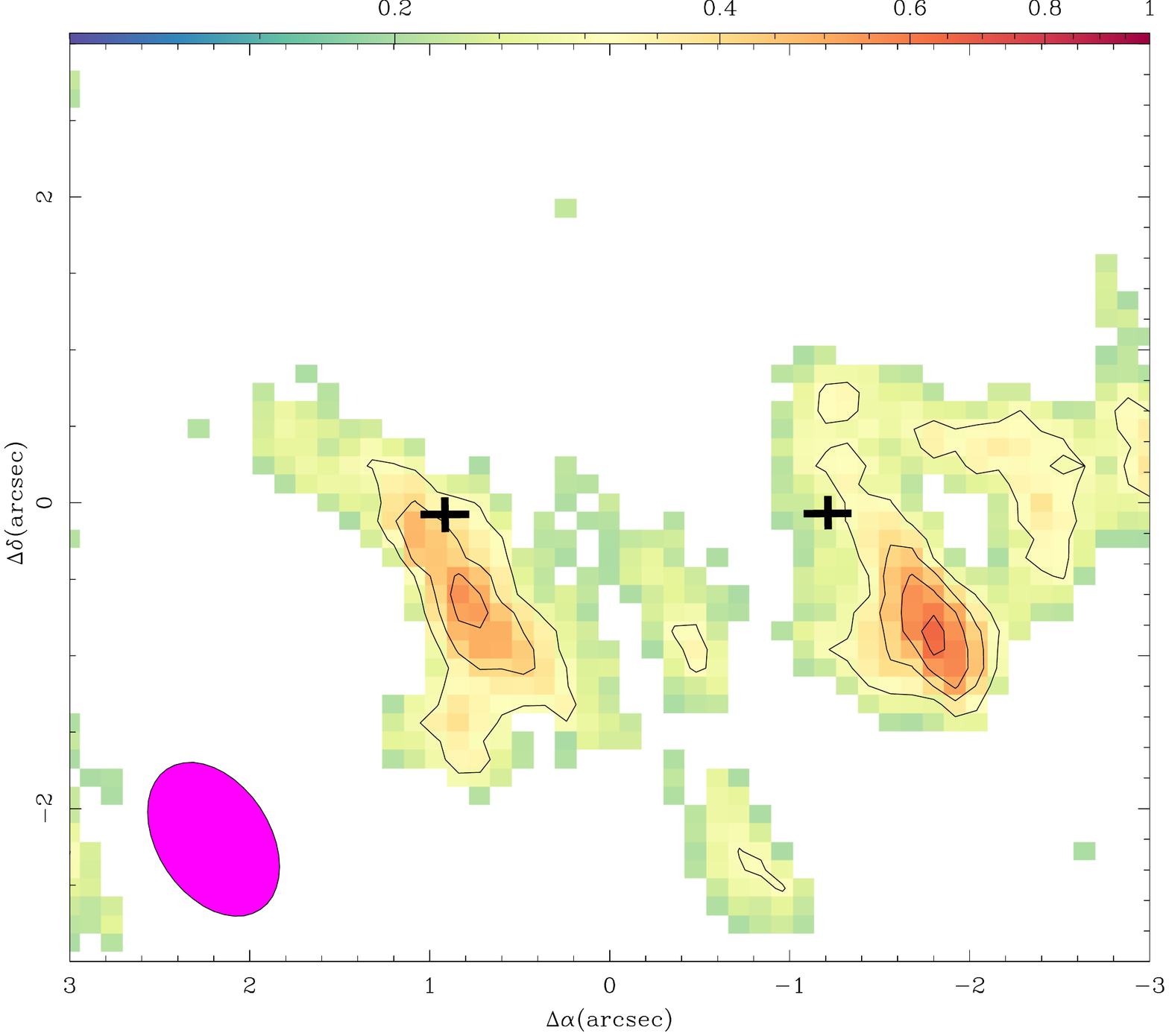}}
\caption{HNCO$(6-5)$ map { over the central 6$'' \times$ 6$''$ region}. Contours are from 3$\sigma$ in increments of 1$\sigma$ where $\sigma$ $\sim$0.1 Jykm/s/beam. {The black crosses indicate the approximate position of the CO$(3-2)$ peak. The beam size is displayed in the bottom left (1$''$.1 $\times$ 0$''$.7, PA=23$^{\circ}$}}
\label{fig:hncomap}
\end{figure*}

\begin{figure}[bhtp]
\centering
\includegraphics[trim={0cm 2cm 0cm 0cm},clip, width=0.9\linewidth, angle=90]{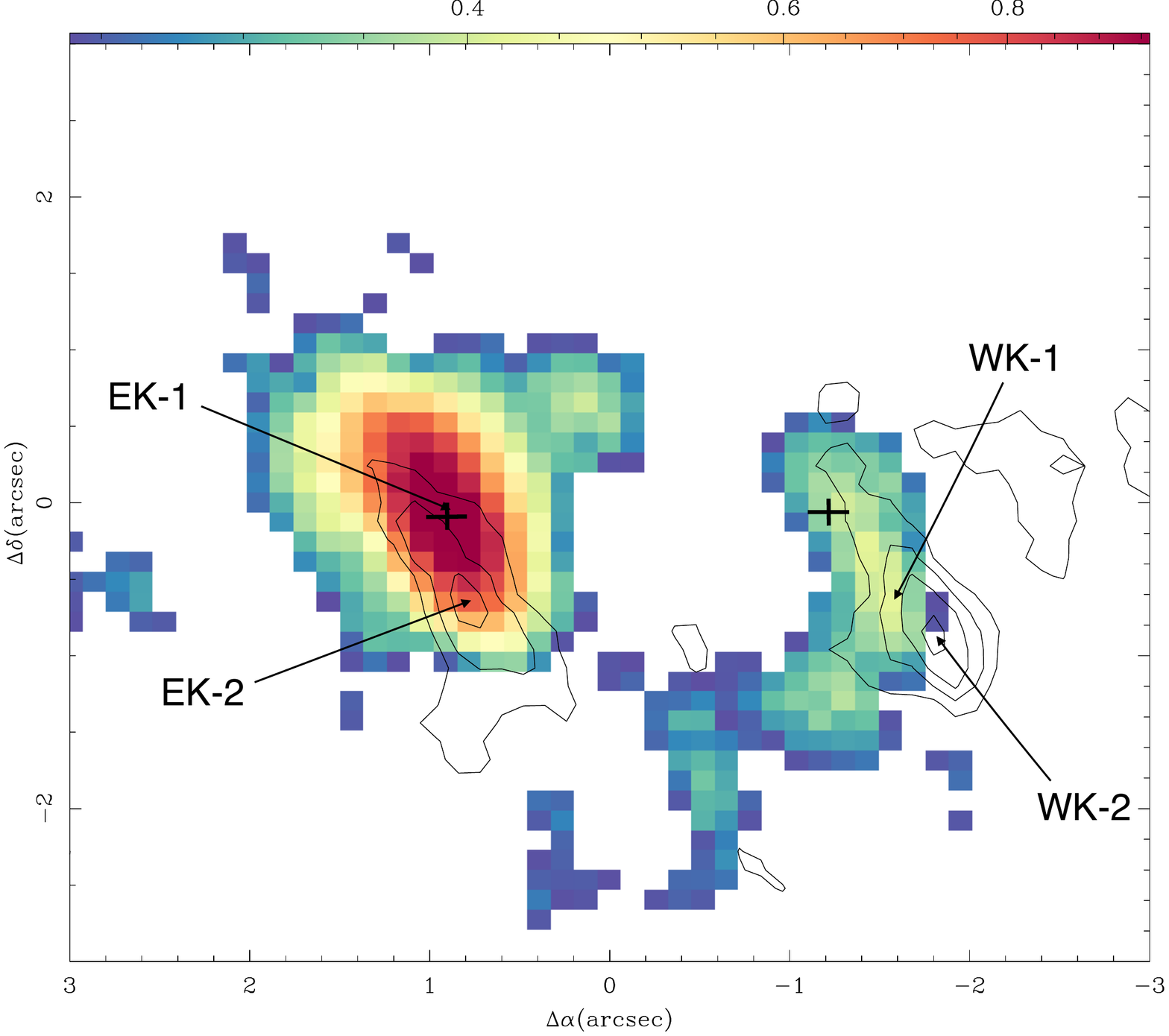}
\caption{{Overlap of SiO (colours) and HNCO (contours). The locations identified for further analysis are labelled. Contours are from 3$\sigma$ in increments of 1$\sigma$ where $\sigma$ $\sim$0.1 Jykm/s/beam. The black crosses indicate the approximate position of the CO$(3-2)$ peak.}}
\label{fig:siohncomap}
\end{figure}

Figure ~\ref{fig:hncomap} shows the HNCO$(6-5)$ map over the same region as the SiO$(3-2)$ map, and integrated across the same spectral window. The strongest emission is seen in the West Knot, but is also clear in the East Knot. There is little to the north or south of the AGN. Nothing is seen around the AGN itself. When compared to both the SiO and CO peaks, it is clear that the HNCO emission is offset in both regions. It is significantly south ($\sim$1$''$) and slightly West of the other peaks. We show this in Figure ~\ref{fig:siohncomap}, where we overlap the two sets of data.

Although we do not split HNCO$(6-5)$ emission into bands due to its weaker emission compared to that SiO$(3-2)$, we analyse HNCO spectra in section ~\ref{spec}.

\begin{figure}[hbtp]
\centering
{\includegraphics[width=1\linewidth, angle=0]{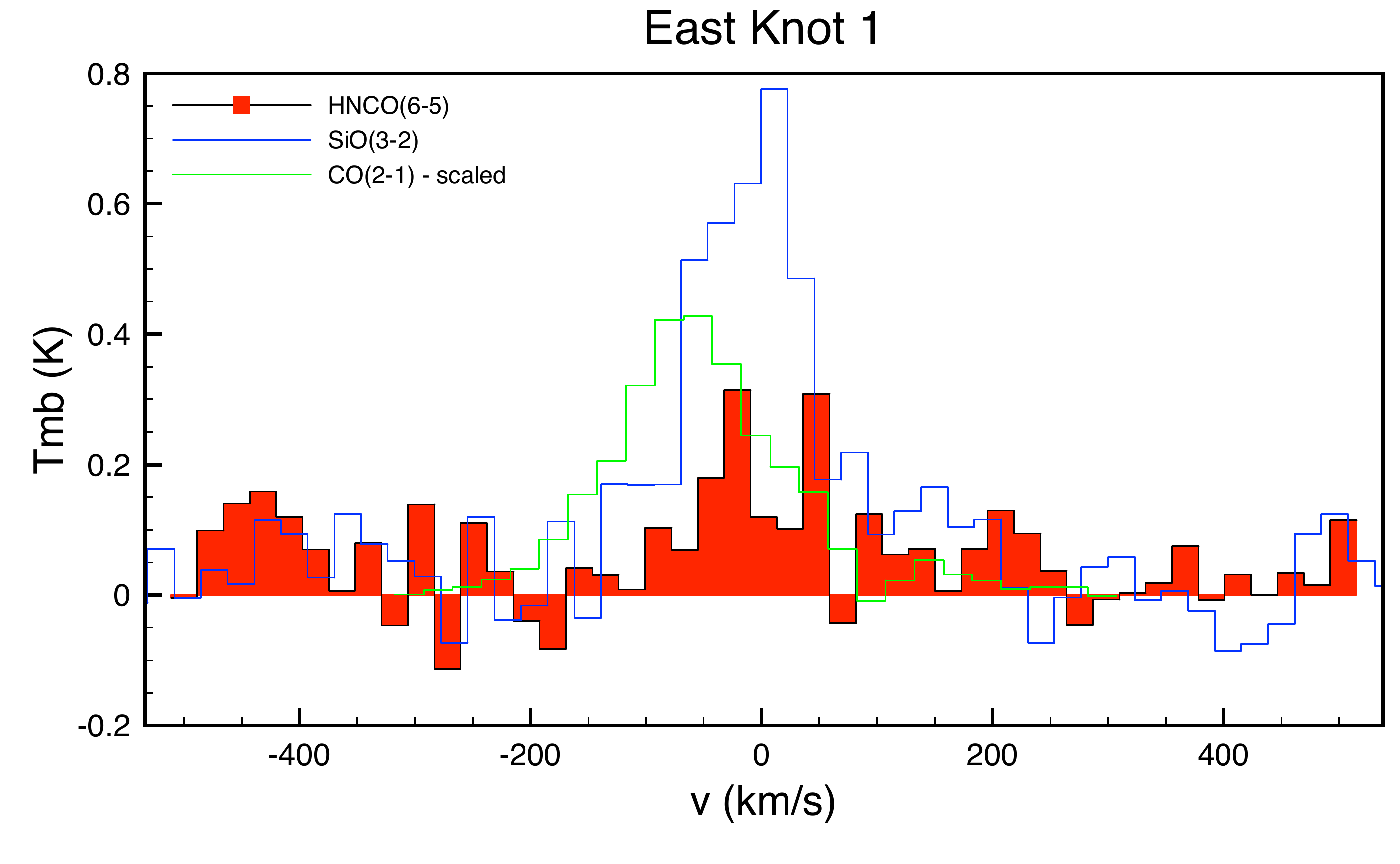}}
{\includegraphics[width=1\linewidth, angle=0]{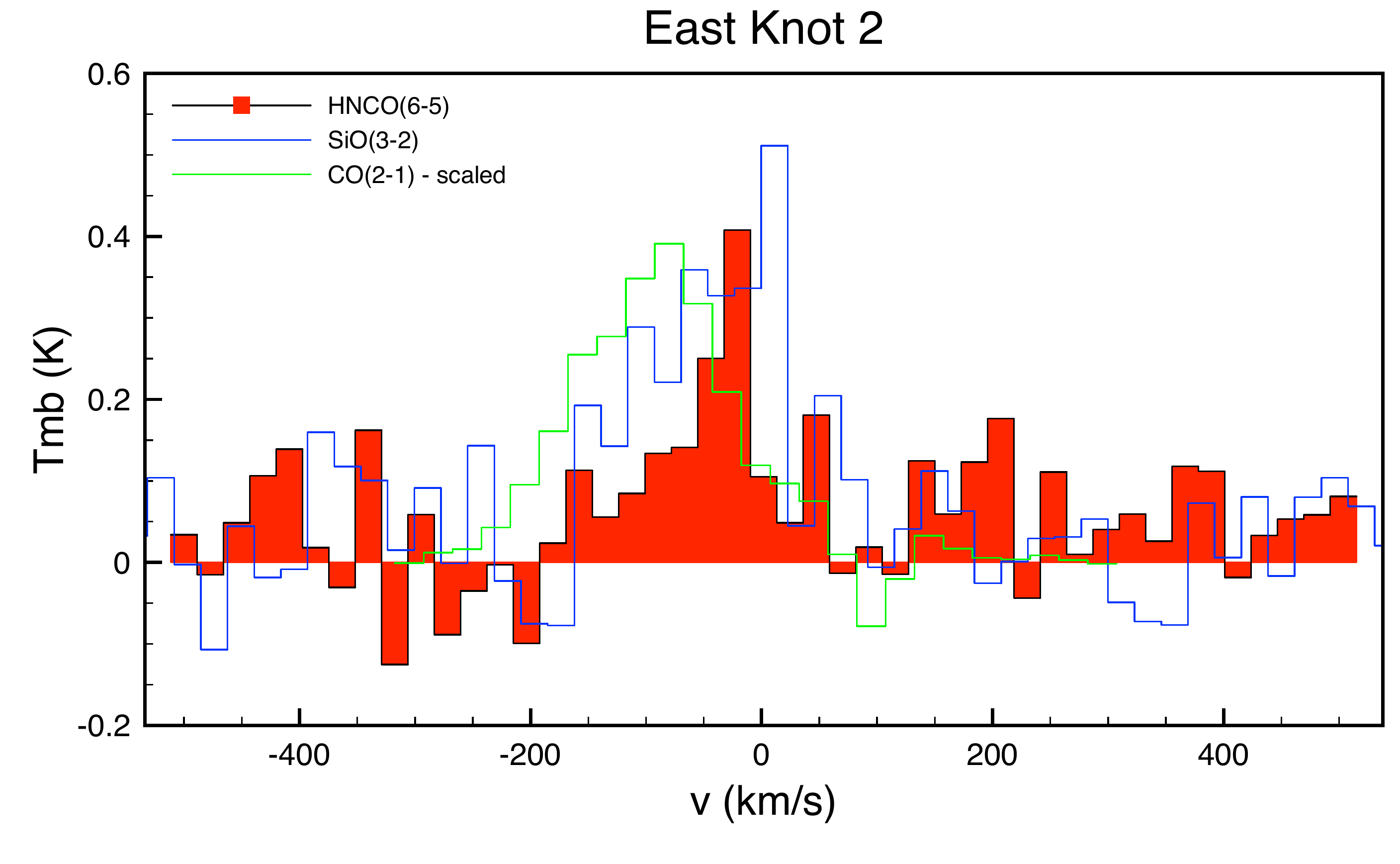}}
{\includegraphics[width=1\linewidth, angle=0]{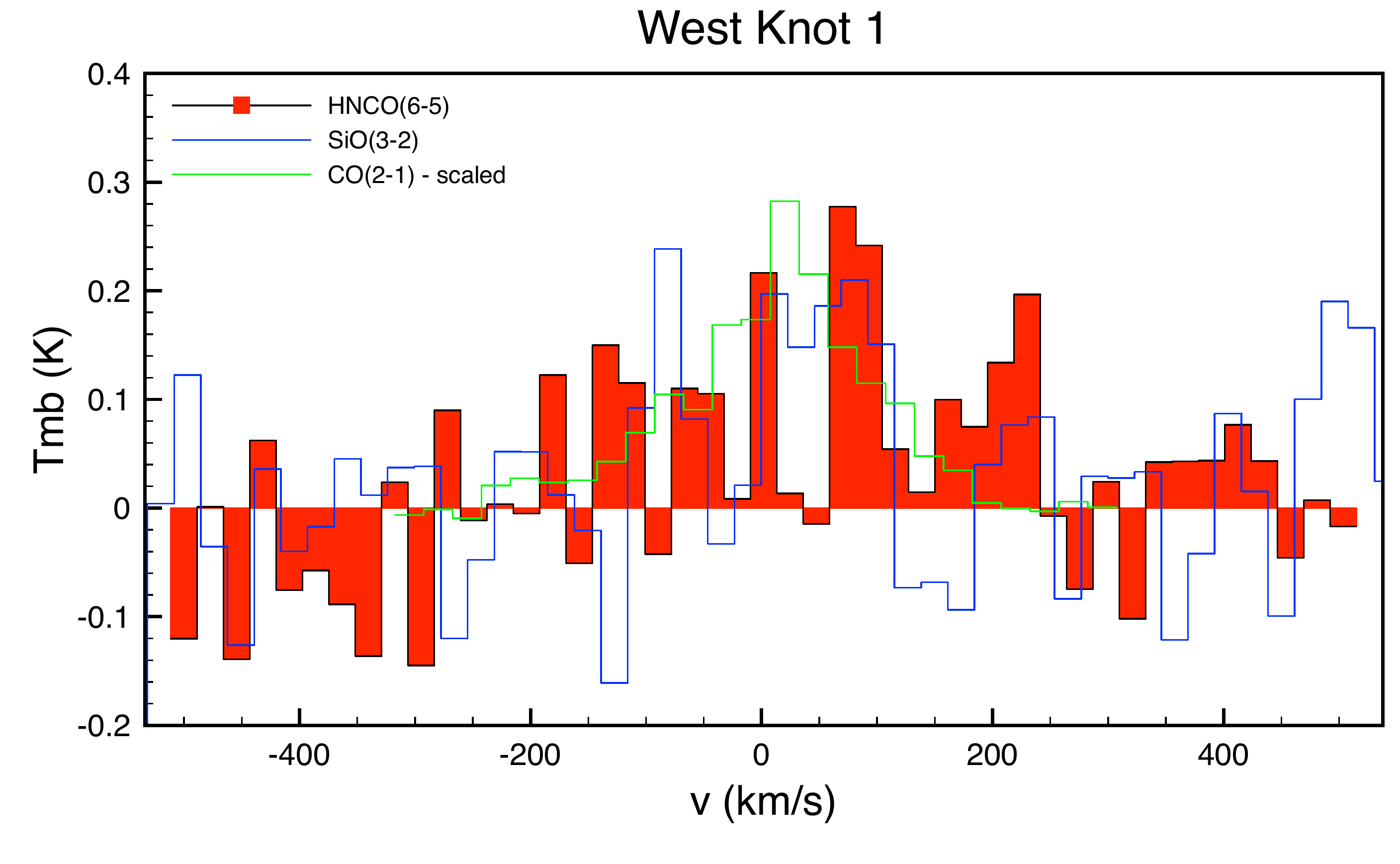}}
{\includegraphics[width=1\linewidth, angle=0]{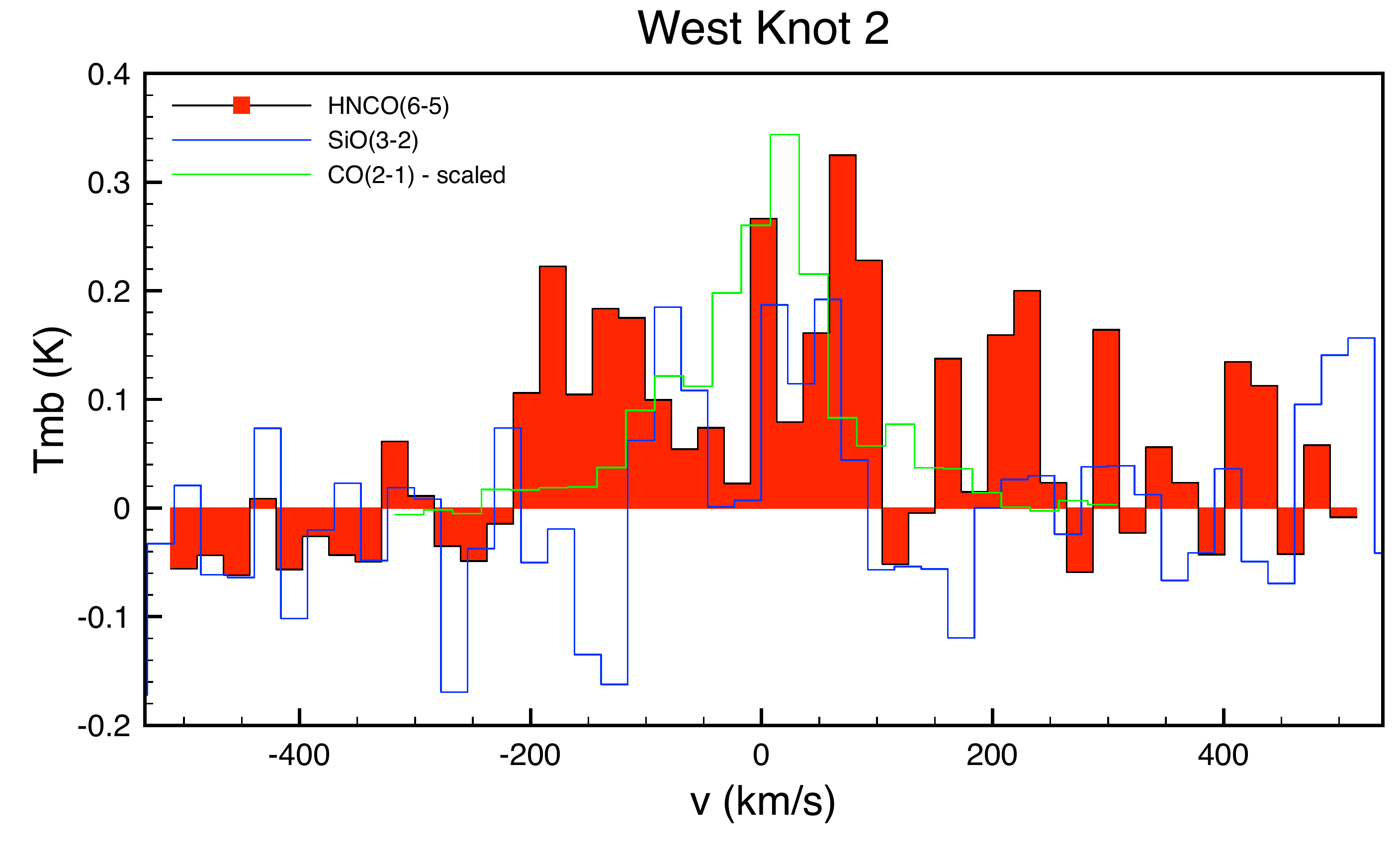}}

\caption{SiO$(3-2)$, HNCO$(6-5)$ and CO$(2-1)$ at four different positions listed in Table ~\ref{tab:positions}. $\Delta$v = 24 km s$^{-1}$. }. 
\label{fig:spectra}
\end{figure}

\section{Spectral analysis} \label{spec}

We extract spectra from four regions of the CND. These are the { SiO peak in the East Knot and West Knot, and the HNCO peak in the East Knot and West Knot}. We extract a spectrum from both species in these regions, giving us eight spectra in total. We also overlay the CO$(2-1)$ line (its peak scaled to the HNCO$(6-5)$ peak), from PdBI observations, degraded to the same resolution as the SiO$(3-2)$ line. We do this to confirm that the HNCO$(6-5)$ emission we see, in particular in West Knot 2, is real. In this location, although the profiles are different, we see contiguous emission from 200 km s$^{-1}$ down to -200 km s$^{-1}$ in CO$(2-1)$. The wings of the CO$(2-1)$ line are at detection confidence level of 4$\sigma$. We also see a clear detection of CO$(3-2)$ (not shown) at high velocities in West Knot 2, from ALMA observations. This is not to suggest that CO is tracing { only shocked gas: CO is present in the shocked region as well as in the other components of the molecular gas}. We use CO only as a guide to see the allowed velocities for SiO and HNCO. As further reassurance our detection of HNCO in the West Knot is real, there are no other detectable transitions of other species near the HNCO$(6-5)$ band frequency.

The four locations are displayed in Figure ~\ref{fig:siohncomap} and Figure ~\ref{fig:spectra}. { In the Figure, the radial velocity is assumed to equal the offset from the galaxy's recession velocity.} We see in East Knot 1, the point of peak SiO emission in the East Knot, a strong SiO line centred around -12 $\pm$ 4 km s$^{-1}$, compared to v$_{sys}$, where there is also evidence of a double peak. In the same location, HNCO displays a similar feature, but is much weaker. It is centred around -6 $\pm$ 15 km s$^{-1}$, compared to v$_{sys}$. Moving to East Knot 2, the Eastern peak of HNCO, we see a notably weaker detection of SiO. Both are centred around $\sim$ 30 km s$^{-1}$, compared to v$_{sys}$. In the position West Knot 1, we see only a marginal increase in SiO signal above the baseline noise, despite this being the peak signal for SiO in the West. We also see weak emission for HNCO in this location. At the HNCO peak, West Knot 2, the SiO spectrum displays the same characteristics as in West Knot 1. This is not surprising judging from Figure ~\ref{fig:siomap}, which shows a diffuse SiO peak. The HNCO peak is weak, but very broad if one assumes a single peak, with a linewidth of 399 $\pm$ 57 km s$^{-1}$. 

\begin{table*}[htbp]
\begin{center}
\caption{Velocity integrated intensities (K km s$^{-1}$) and ratios}
\label{tab:ratiovel}
\begin{tabular}{c c c c}
Location & HNCO$(6-5)$ & SiO$(3-2)$ & HNCO$(6-5)$ / SiO$(3-2)$ \\
\hline
EK-1 & 20 $\pm$ 8 & 63 $\pm$ 4 & 0.31 $\pm$ 0.13   \\
EK-2 &  22 $\pm$ 7  & 56 $\pm$ 5 & 0.39 $\pm$ 0.12  \\
WK-1 & 41 $\pm$ 7 & 22 $\pm$ 4 & 1.8  $\pm$ 0.46 \\
WK-2 & 60 $\pm$ 8 & 17 $\pm$ 5 & 3.5  $\pm$ 1.2 \\
\end{tabular}
\end{center}
\end{table*}

%

\begin{figure}[htbp]
\centering
\includegraphics[width=0.9\linewidth, angle=0]{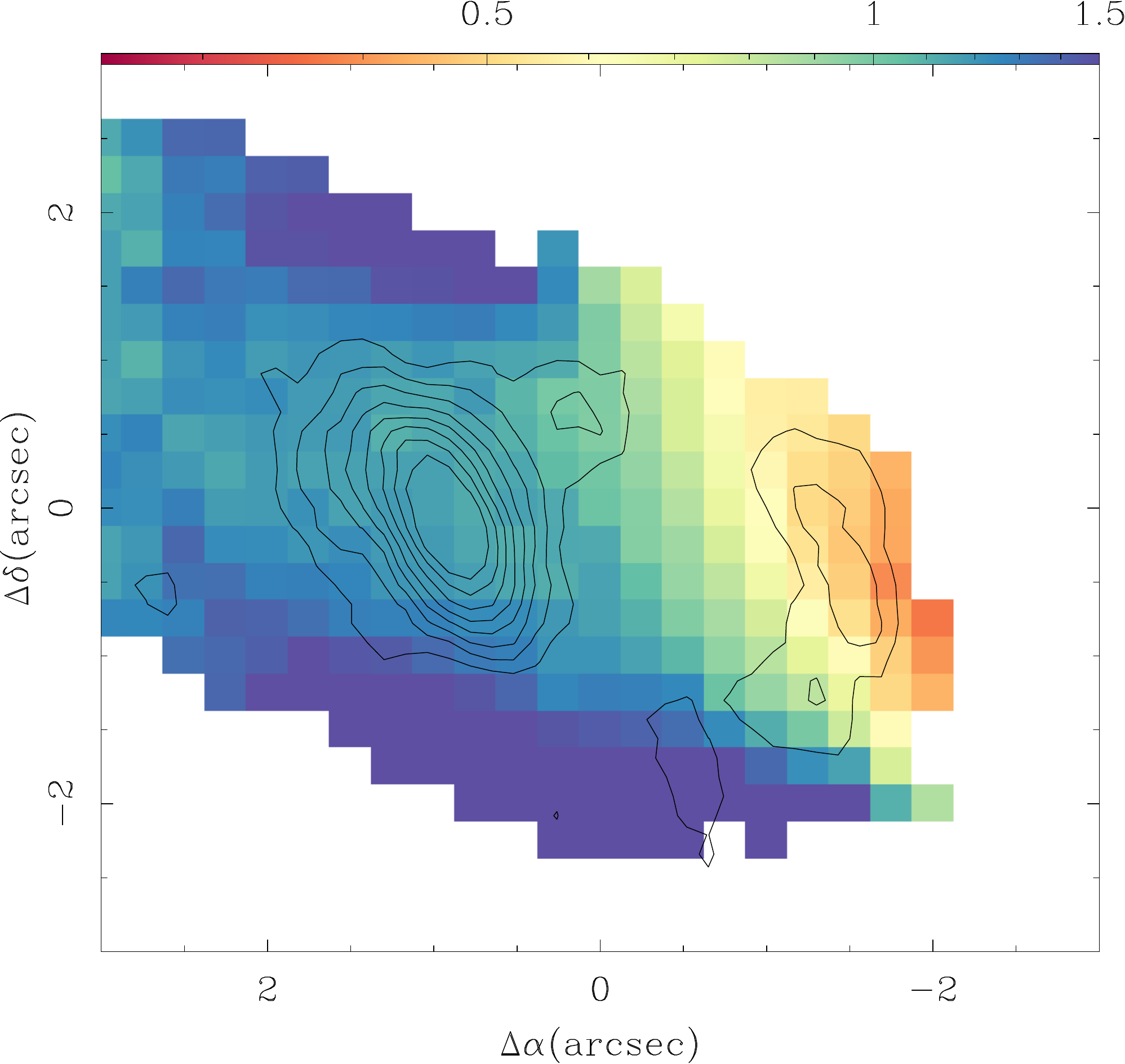}
\caption{Colours: Velocity integrated intensity ratio of SiO$(3-2)$/SiO$(2-1)$ at the resolution of the SiO$(2-1)$ line (3$''$.6 $\times$ 2$''$.1, PA=29$^{\circ}$). Contours: SiO$(3-2)$ at original resolution (1$''$.1 $\times$ 0$''$.7, PA=23$^{\circ}$).}
\label{fig:ratiomap}
\end{figure}

We compare the line { intensities ratios for each location (in T$_{mb}$ $\Delta v$ units) . The intensities and ratios are shown in  Table ~\ref{tab:ratiovel}}. In the East, locations 1 and 2, SiO dominates by a factor of 3. In the West, West Knot 1, HNCO is a factor of 1.8 stronger, and twice that in West Knot 2. As a first approximation, the HNCO / SiO ratio may be considered a measure of the shock strength, since it takes a stronger shock event to increase the SiO abundance. This indicates that if we take the velocity integrated line { intensity} ratios, shocks in the East Knot are significantly stronger than those in the West Knot. The average ratio in the East Knot is HNCO$(6-5)$ / SiO$(3-2)$ = 0.35 $\pm$ 0.13 and in the West Knot, HNCO$(6-5)$ / SiO$(3-2)$ = 2.5 $\pm$ 1.3. In addition, the shock strength might be considered to be relatively constant to the East, but due to the disparity in ratio in the two locations in the West Knot, shock velocities may differ on comparatively small scale.

In addition, using SiO$(2-1)$ data from \citet{2010A&A...519A...2G}, we produce a SiO$(3-2)$/SiO$(2-1)$ ratio map. To do this the resolution of the SiO$(3-2)$ observations is degraded to match the lower transition. The map is displayed in Figure ~\ref{fig:ratiomap}. There is a clear gradient from East to West. In the East the ratio is $\sim$ 1.2 but this drops to $\sim$ 0.5 in the West. This compares well with previous observations in \citet{2014A&A...567A.125G} and \citet{2014A&A...570A..28V}, who find a comparatively higher excitation of other molecular lines in the East Knot compared with the West Knot. 

Our velocity integrated line { intensity ratio (in T$_{mb}$ $\Delta v$ units)} (HNCO$(6-5)$ / SiO$(3-2)$) values in the East Knot are comparable to the inner disk of NGC 253, which were calculated using the slightly lower J transitions, HNCO$(5-4)$ / SiO$(2-1)$ \citep{2015ApJ...801...63M}. The NGC 253 compact (r $\approx$ 170 pc) region has a large quantity of dense gas and an increased rate of star formation. The values are also comparable to M82 \citep{2009ApJ...706.1323M}. The outer disk of NGC 253 shows similar ratios to that of the West Knot. The outer disk is suggested to be a region where gas is flowing radially inwards\citep{2000A&A...355..499G}, in contrast to the outflowing gas in NGC 1068. They also match well with the shocked Giant Molecular Clouds (GMCs) in the nearby spiral barred galaxy Maffei 2 \citep{2012ApJ...755..104M}.

We note that we can not however simply correlate the shock strength with the HNCO/SiO ratio, especially as several gas phase reactions may contribute to the formation and destruction of both molecules. For example, HNCO is readily photodissociated at a rate $\sim$ 30 times faster that SiO \citep{1995ApJS...99..565S}. If in the presence of significant UV radiation, HNCO abundance will be significantly lowered compared to SiO. We would therefore see weaker emission and our ratios would be decreased by this effect.

\section{Analysis}

In this section we attempt to quantify the differences we observe between SiO $(3-2)$ and HNCO $(6-5)$ emission by position in the CND.
There is strong SiO $(3-2)$ emission in the East Knot, peaking at East Knot 1, whereas HNCO $(6-5)$ emission is relatively weak in the same region. Conversely, we see that SiO $(3-2)$ and HNCO $(6-5)$ emission is more equal in the West Knot, and HNCO $(6-5)$ emission is stronger than SiO $(3-2)$ in West Knot 2. In order to explain these differences, we complete a three-phase analysis: a basic LTE analysis, a further radiative transfer modelling using RADEX, and a chemical modelling using UCL\_CHEM { (Viti et al. 2004; Viti et al. 2011)}, along the lines of the procedure set in Viti et al. (2014).

\subsection{LTE}

Here we calculate column densities, $N$, of SiO and HNCO, assuming local thermodynamic equilibrium (LTE) and optically thin emission. For that we use equation 1.

\begin{equation}
N_{mol} = \num{1.67E14} \frac{Q(T_{rot})}{\mu^2\nu S} exp \bigg(\frac{E_u}{kT_{rot}}\bigg) \int T_{MB} dv 
\end{equation}

Where {\it Q(T$_{rot}$)} is the partition function, $\mu$ is the dipole moment in Debye, {\it S} is the line strength. $E_u$ is the upper energy level, $T_{rot}$ is the rotational temperature of the molecule. The units of the integral are K km s$^{-1}$; we therefore convert our data from Jy/beam to $T_{MB}$. \citet{2014A&A...570A..28V} find { that the gas must have a kinetic temperature of at least 
$\sim$ 50 K}. We therefore use this temperature as our first T$_{rot}$ in order to get the values of $N$(SiO) and $N$(HNCO) for each location. It is possible the average temperature of the gas is \textgreater 50 K, especially if the gas is shocked. To account for this, we also calculate column densities for 100 K and 200 K. These temperatures also correlate with the findings of \citet{2014A&A...570A..28V}. The calculated column densities are displayed in Table ~\ref{tab:colden}. HNCO $(6-5)$ is a band of several blended lines in our observations. To calculate a column density, we use average values from the strongest band, HNCO$(K_{\rm p},K_{\rm o})$)=(0,6)-(0,5). {It is possible that the gas density may be below the critical density of the transitions. However, since there may be shocks present, the density of the gas traced by these molecules may be much higher than the gas surrounding it. The critical densities, n$_{cr}$, of SiO$(3-2)$ and the strongest band of HNCO$(6-5)$ are in the order of \num{e6} cm$^{-3}$. Since the gas may not be thermalised, our LTE column density values should be used with caution.}

\begin{table}[htbp]
\begin{center}
\caption{LTE column densities in each location}
\label{tab:colden}
\begin{tabular}{c c c}
Location &  $N$(SiO) (cm$^{-2}$) & $N$(HNCO) (cm$^{-2}$)   \\
\hline
\multicolumn{3}{c}{T = 50 K} \\
EK-1 & \num{2.4e14} & \num{3.2e14} \\
EK-2 & \num{1.5e14} & \num{6.8e14} \\
WK-1 & \num{6.2e13} & \num{8.4e14} \\
WK-2 & \num{4.9e13} & \num{1.3e15} \\
\hline
\multicolumn{3}{c}{T = 100 K} \\
EK-1 & \num{4.3E+14} & \num{7.2E+14} \\
EK-2 & \num{2.6E+14} & \num{1.5E+15} \\
WK-1 & \num{1.1E+14} & \num{1.9E+15} \\
WK-2 & \num{8.6E+13} & \num{2.9E+15} \\
\hline
\multicolumn{3}{c}{T = 200 K} \\
EK-1 & \num{8.0E+14} & \num{2.0E+15} \\
EK-2 & \num{4.8E+14} & \num{4.2E+15} \\
WK-1 & \num{2.1E+14} & \num{5.2E+15} \\
WK-2 & \num{1.6E+14} & \num{7.8E+15} \\
\end{tabular}
\end{center}
\end{table}

Comparing the column densities from Table ~\ref{tab:colden}, we notice that, in East Knot 1, where SiO emission peaks, column densities for SiO and HNCO are quite similar, but $N$(HNCO) is greater than $N$(SiO) at all temperatures. In East Knot 2, the peak of HNCO emission in the East Knot, $N$(HNCO) is half an order of magnitude higher than $N$(SiO). In the West Knot, $N$(HNCO) is between 1-2 orders of magnitude greater than $N$(SiO). This is assuming that SiO and HNCO emission is coming from the same gas component. However, under the assumption that SiO traces a fast shock and HNCO traces slower shocks, the gas emitting in SiO may well be warmer. We find that, if one assumes the gas emitting SiO and HNCO has the same temperature, then East Knot 2, and West Knot 1\&2, still show that HNCO column density is higher than that of SiO. In East Knot 1, if the temperature of the SiO emitting gas is 100 K, and the HNCO emitting gas is 50 K, $N$(SiO) is greater than $N$(HNCO). If $N$(SiO) is then calculated assuming the temperature is 200 K, it becomes greater than $N$(HNCO) by a factor of $\sim$ 2.5. 

\subsection{RADEX} \label{RADEX}

In order to further characterize the emitting gas in SiO and HNCO, we run a RADEX \citep{2007A&A...468..627V} analysis. Since we only have one transition for each molecule, we also take the SiO$(2-1)$ line from \citet{2010A&A...519A...2G} and the HNCO$(5-4)$ line from \citet{2014PASJ...66...75T}. We complete a separate analysis for each location, using the brightness temperature of the four observed lines, {therefore assuming a common filling factor}. Our grid of models are run varying hydrogen number density from \num{e3} cm$^{-3}$ to \num{e8} cm$^{-3}$ and temperature from 10 K to 300 K. We use an initial input of column density based on our LTE calculations at 50 K. We then vary the column density by up to an order of magnitude. For the purpose of our analysis, we divide our model grid in subsets where we consider a set of models as models with a constant column density. This results in sets containing over 15,000 models. {RADEX requires an input value for column density over line width (N$_{mol}$/$\Delta$V). We measure the line width by fitting a Gaussian to each line. The mean value of the line width is $\sim$ 150 km s$^{-1}$. This is the value we use for all our RADEX models. We vary this value by a factor of 2 as a check of its influence on our results. A smaller or larger line width has a marginal effect on our best fit parameters, but not enough to alter our conclusions.}

We calculate a reduced $\chi^{2}$ for each model compared to observations in each of our four locations. The reduced $\chi^{2}$ equation used is the same as in \citet{2014A&A...570A..28V}, although here we compare observed and modelled brightness temperature, not ratios:

\begin{equation}
\chi^{2}_{\rm red} = \frac{\rm1}{\rm K} \sum_{n=1}^{\rm K} [\rm log(T_{o}) - \rm log(T{_m})]^{2}/\sigma^{2}
\end{equation}

\begin{figure*}[hbtp]
\centering
{\includegraphics[width=0.4\linewidth, angle=0]{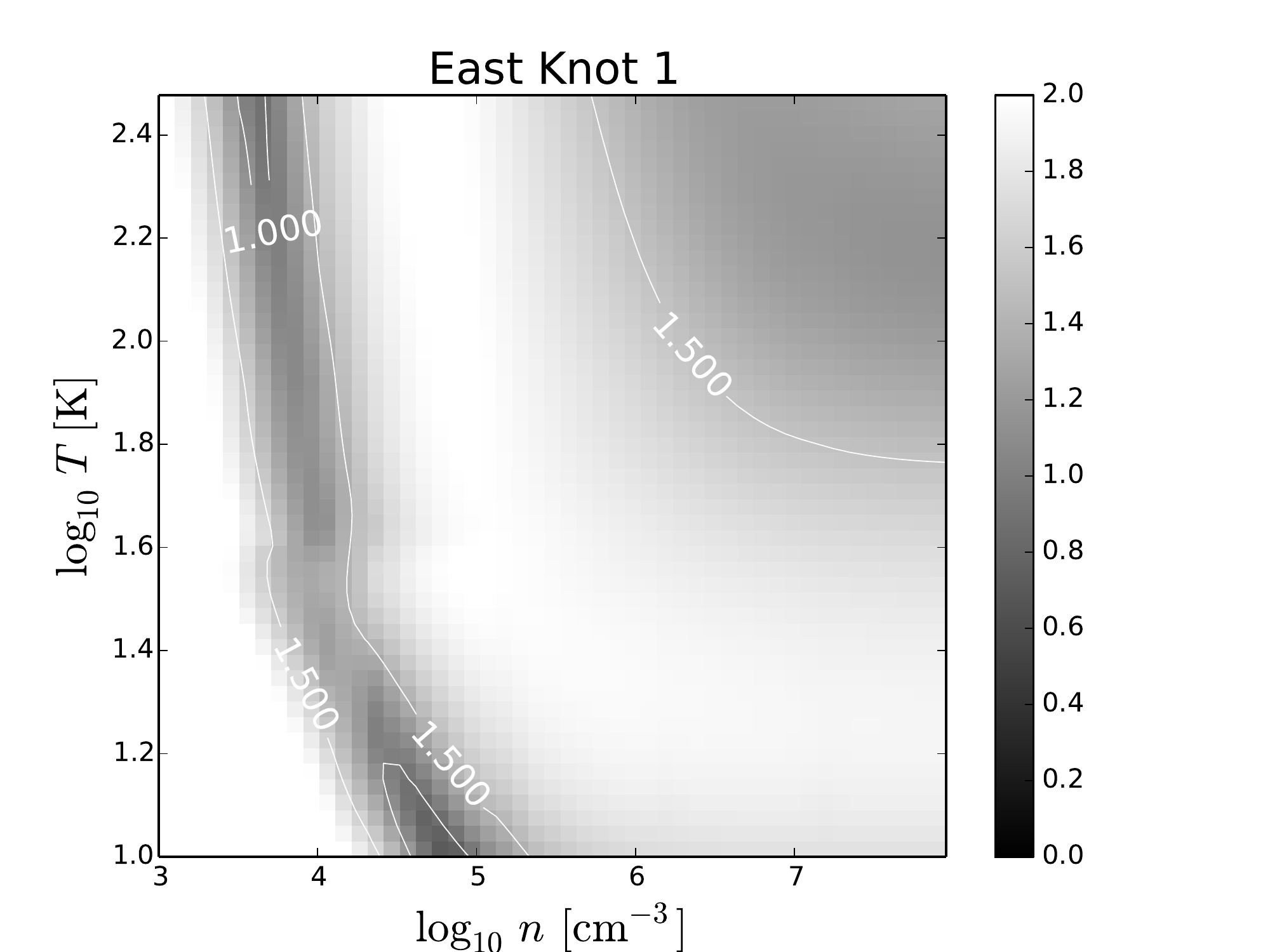}}
{\includegraphics[width=0.4\linewidth, angle=0]{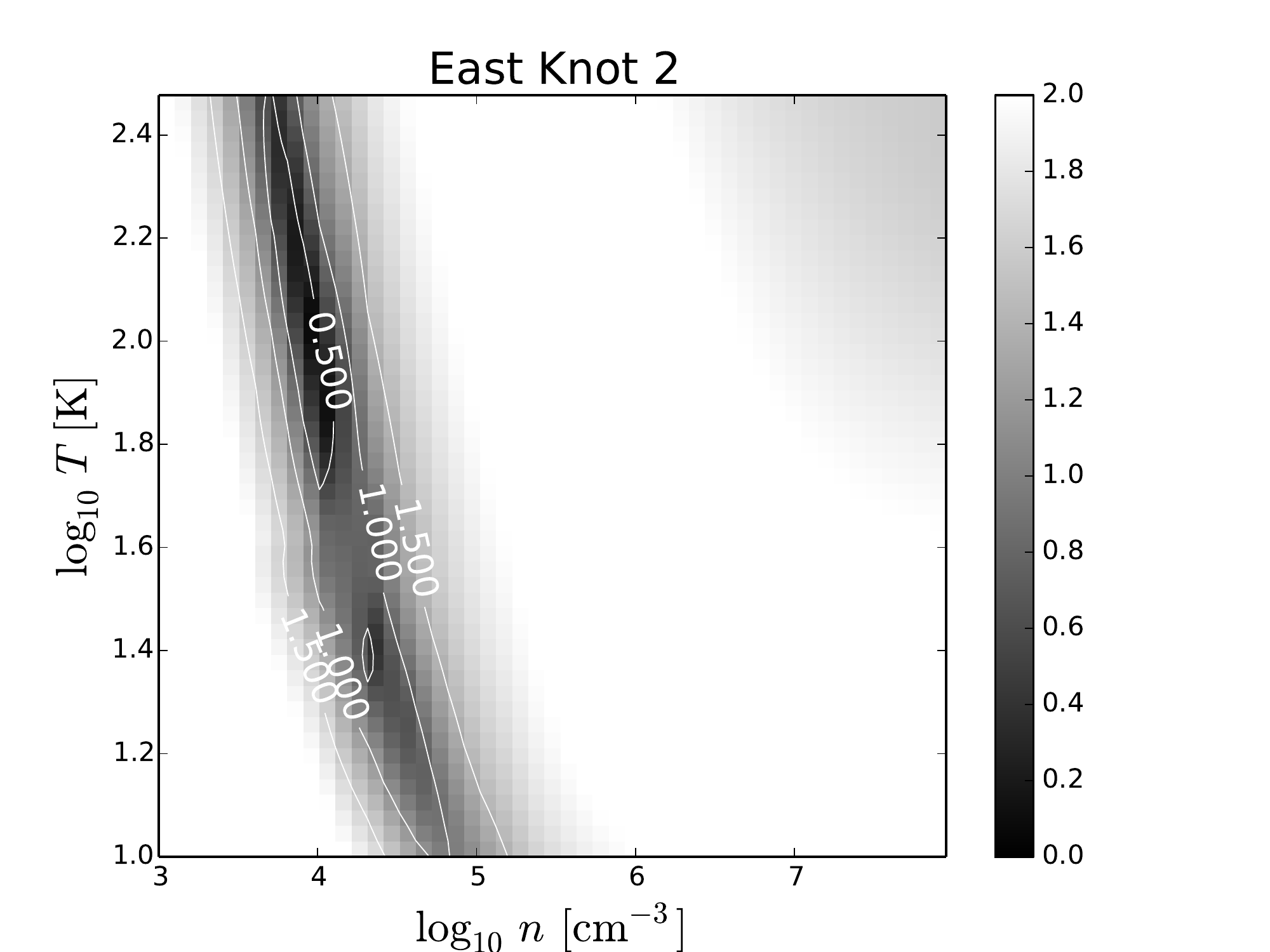}}
{\includegraphics[width=0.4\linewidth, angle=0]{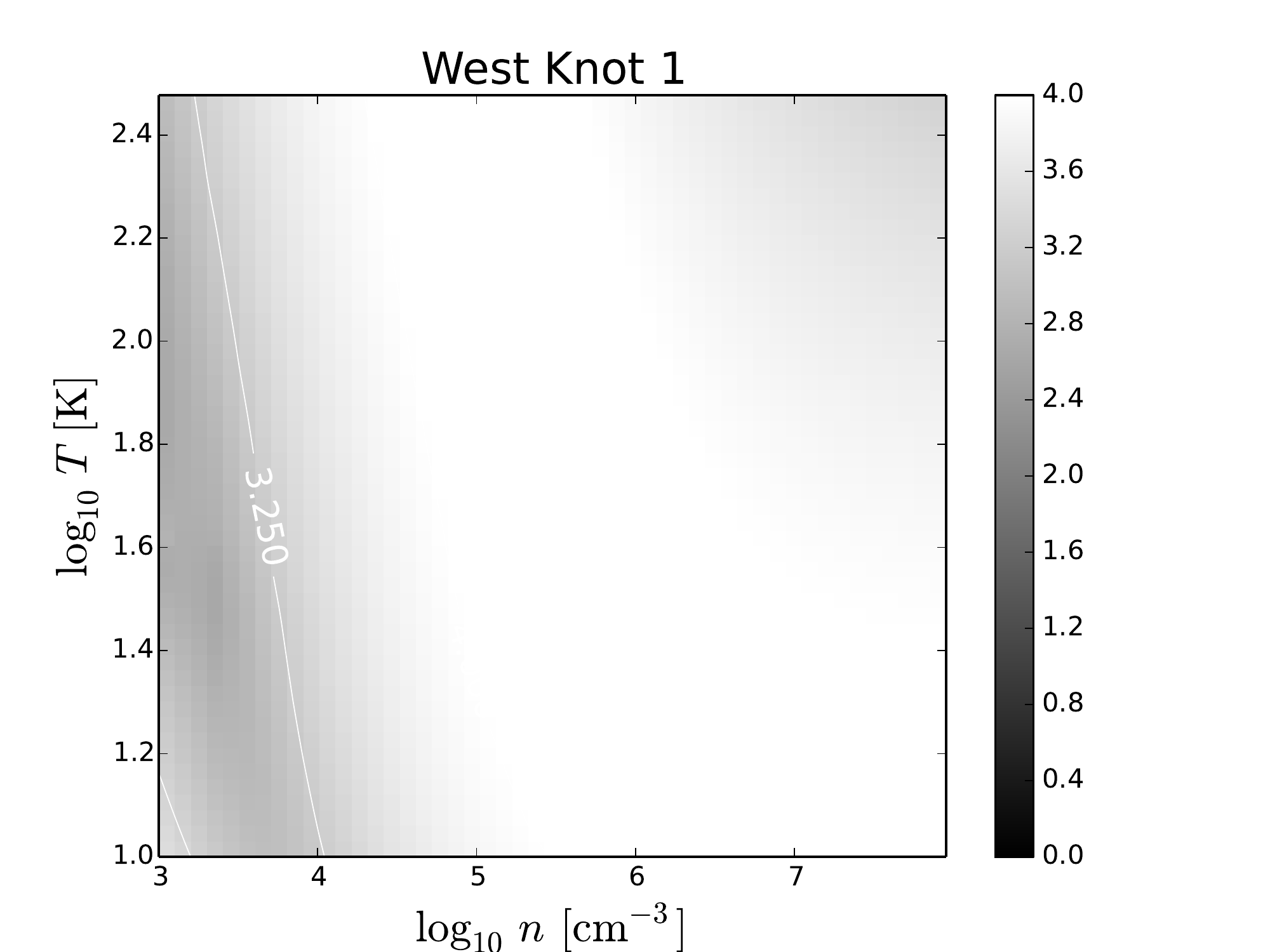}}
{\includegraphics[width=0.4\linewidth, angle=0]{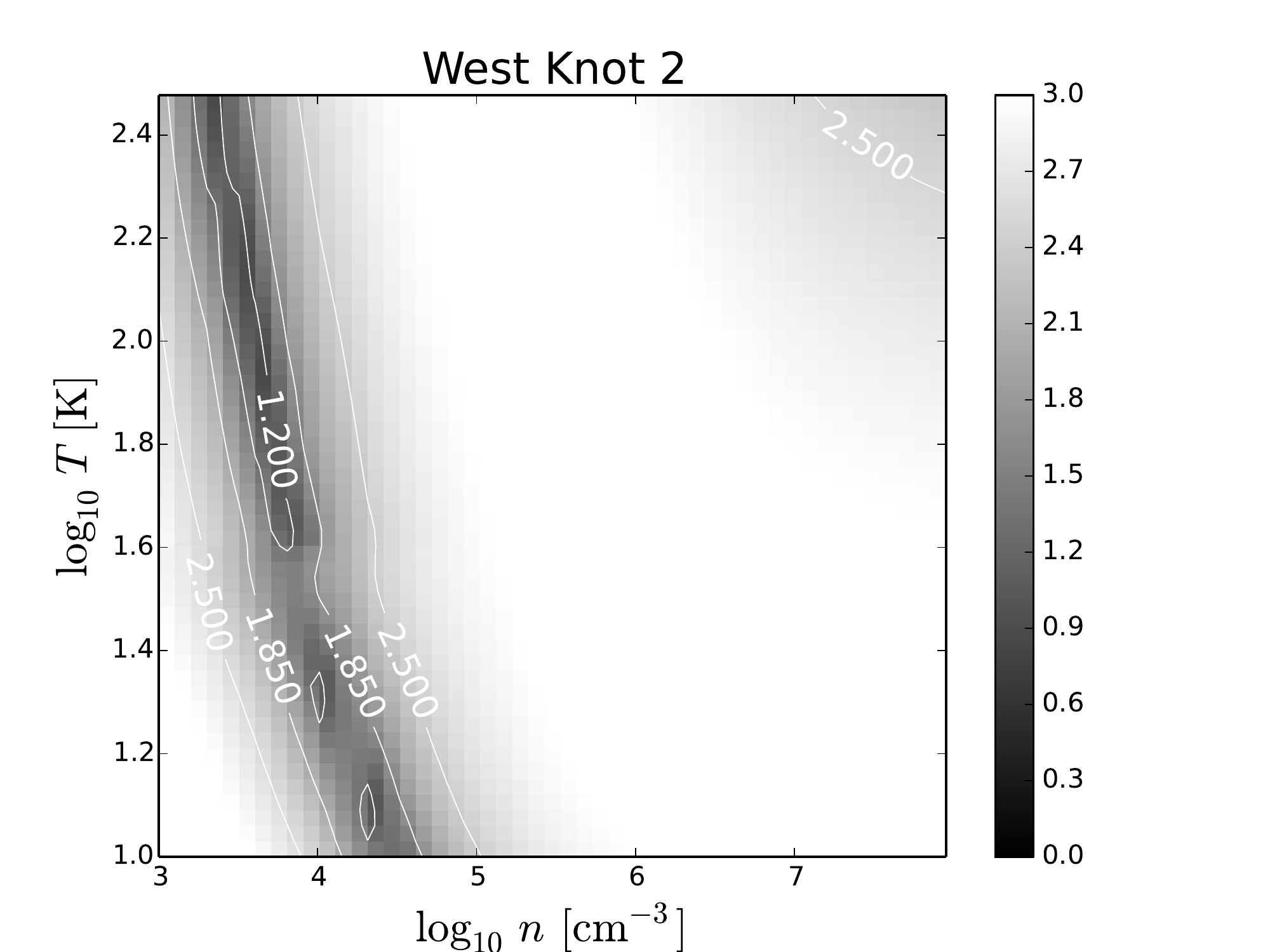}}
\caption{Log of $\chi^{2}$ fits from RADEX modelling for each location {with varying temperature and density. Darker regions show a lower $\chi^{2}$ and therefore better fit than lighter regions}}
\label{fig:chi2both}
\end{figure*}

Where K is defined as N - n, where N is the number of observed lines and n is the number of varied parameters. In this case, N = 4 and n = 2. $\rm T_{o}$ and $\rm T_{m}$ are observed and modelled brightness temperature respectively, and $\sigma$ is the uncertainty on the observed brightness temperature. We estimate $\sigma$ to be 20\% for all our observations based on the systematic uncertainty from calibration. Our results are displayed in Figure ~\ref{fig:chi2both}. We find that the lowest median $\chi^{2}$ values are for models within the set ran with our initial LTE estimated column density for 50 K, although the differences are quite small. In this section we therefore concentrate on analysis of this set, but note that the results of using a higher column density would lead to similar conclusions.

We describe below our findings per each location: \\ 
\begin{itemize}
\item East Knot 1: the values of $\chi^{2}$ are generally quite high. Nevertheless the lowest $\chi^{2}$ constrains both density and temperature at around 10$^5$--10$^6$ cm$^{-3}$ and 10--30 K. However we also note that if one accepts log of $\chi^{2}$ of just 0.2 higher than the best fit then densities may be as high as 10$^7$ cm$^{-3}$ and temperatures between 40-65 K. 

\item East Knot 2: we obtain a fairly low log of $\chi^{2}$ for a density between 10$^4$ and 10$^5$ cm$^{-3}$ but temperature is quite poorly constrained as low $\chi^{2}$ values are found across our entire temperature range.

\item West Knot 1: this location is not well fit at all. The lowest (but still quite high) $\chi^{2}$ are found for either a combination of very low densities ($<$ 10$^4$ cm$^{-3}$) and a large range of temperatures or very high densities and temperatures ($>$ 10$^7$ cm$^{-3}$ and $>$ 250K respectively).

\item West Knot 2: this location is well fit by a density between 10$^4$--10$^5$ cm$^{-3}$; however we are not able to constrain a temperature. 
\end{itemize}

While we were not able to find a convincing solution for West Knot 1, 
two main results for the remaining Locations emerged: (i) the density of the gas is at least 10$^4$ cm$^{-3}$, with East Knot 1 having the highest density and East Knot 2 and West Knot 2 having a similar density (consistent with all previous work on dense gas from \citealt{2011ApJ...736...37K} and \citealt{2014A&A...570A..28V}); (ii) the temperature of the gas is generally not well constrained. {This may be due to the upper energy levels of our transitions being significantly lower than the actual temperature of the gas. However, \citet{2014A&A...570A..28V}, using mostly transitions with upper energy levels similar to that of HNCO$(6-5)$, conclude that the temperature of the gas in the West Knot may be as high as 200 K. For HNCO$(6-5)$, E$_{u}$ $\approx$ 65 K. Alternatively, it also may be an indication that the gas component(s) where SiO and HNCO emission originates from are not at a constant temperature. This is not surprising since the emission may be coming from a shocked region, where temperatures vary swiftly with time and space (and the cooling rate is proportional to the square of the density). We investigate the shock chemistry that may give rise to the SiO and HNCO emissions in the next Section.}

{ We note that} our RADEX modelling findings are very similar to what has been found previously in \citet{2014A&A...570A..28V}. However we are much more limited by the number of transitions we have. We should also consider that there could be an uncertainly in our best fit parameters as high as a factor of 10 or more \citep{2016ApJ...819..161T}.

\subsection{Chemical modelling}

Following the methodology of Viti et al. (2014) we now adopt a chemical model in order to determine the origin of the emission in HNCO and SiO as well as shed light on the temperature of the gas. In particular, we investigate whether the passage of 
shocks may significantly affect the production or destruction of SiO or HNCO in the gas. We use UCL\_CHEM, a time-dependent gas-grain chemical model \citep{2004MNRAS.350.1029V} coupled with a parametric shock code \citep{2008A&A...482..549J}. { We note that Viti et al. (2014) did not model the gas of the CND using the coupled version of UCL\_CHEM: their grid of models did include some models where a shock was simulated by simply increasing quickly the temperature of the gas, which then remained at high temperatures ($\sim$ 1000 K) for a short period of time before cooling down to either a temperature of 100 or 200~K. In this work we aim at a detailed modelling of the shock process and the chemical model we use is therefore coupled with a parametric shock code.} Details of the coupled code can be found in \citet{2011ApJ...740L...3V}. The model is ran in two phases. Phase I simulates the gas phase chemistry during the formation of high or medium density clumps, along with a freeze out of gas phase species onto dust grain mantles and possible subsequent surface reactions. Other than molecular hydrogen, all initial species are atomic. Initial abundances were set to be solar, other than Si, where we vary an amount that is depleted into dust grain nuclei: this parameter is of particular importance when discerning between shock and non-shock models because in the former the Si in the nuclei would be sputtered back into the gas phase. Phase II simulates either the passage of the shock or simply an increase in gas and dust temperature due to energetics events such as a starburst; it then follows the time evolution of the chemistry in the gas and on the grain mantles. For the shock models this phase includes a plane-parallel shock component with a set velocity, V$_{s}$, temperature, T$_{max}$ and saturation time, t$_{sat}$. For both shock and non shock models, the grain mantle, where both HNCO and Si are present, is sublimated; in addition in the shock models the nuclei of the dust grains may be sputtered, depending on the shock velocity. 

\begin{figure*}[htbp]
\centering
\includegraphics[width=0.3\linewidth, angle=0]{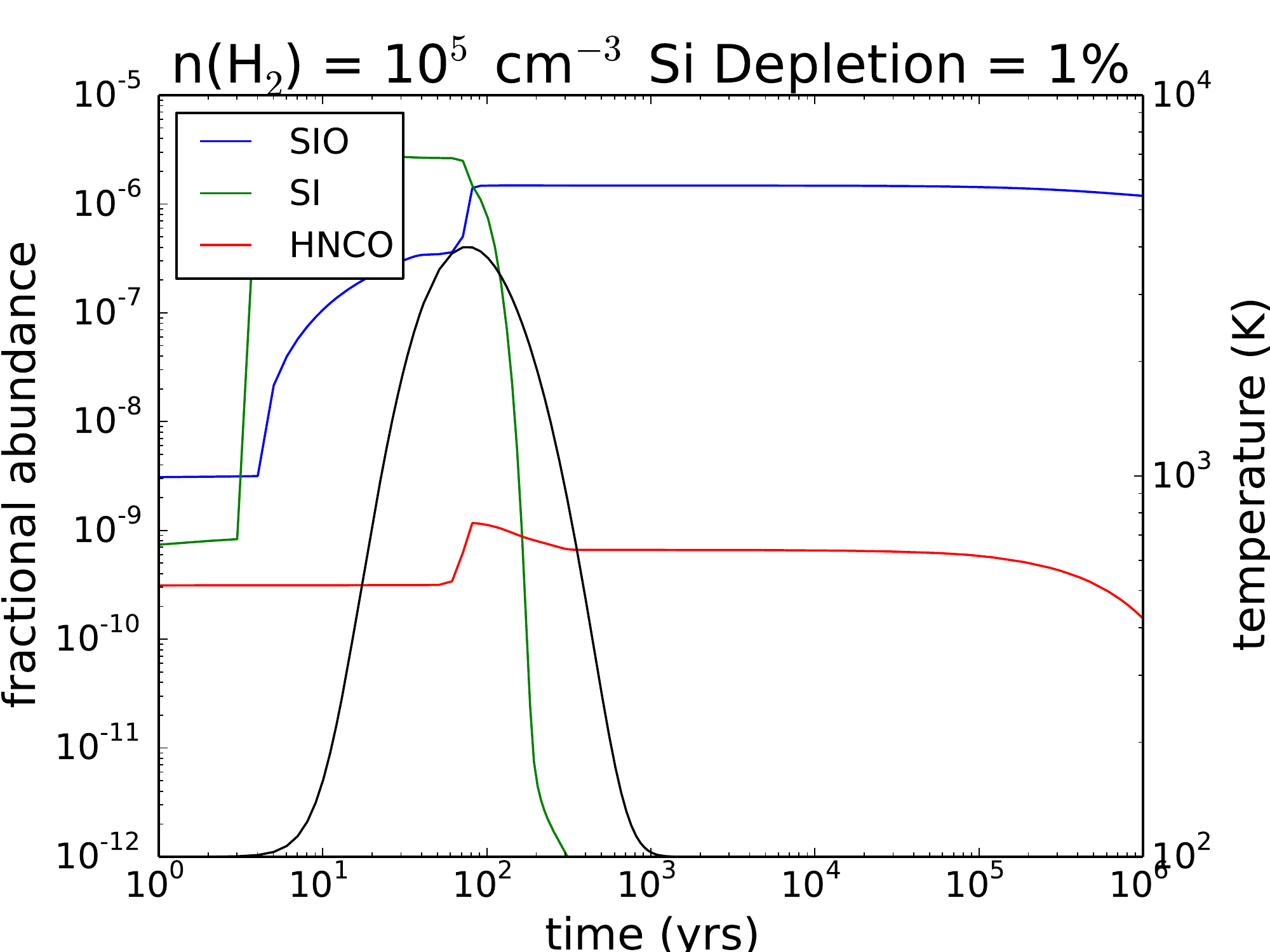}
\includegraphics[width=0.3\linewidth, angle=0]{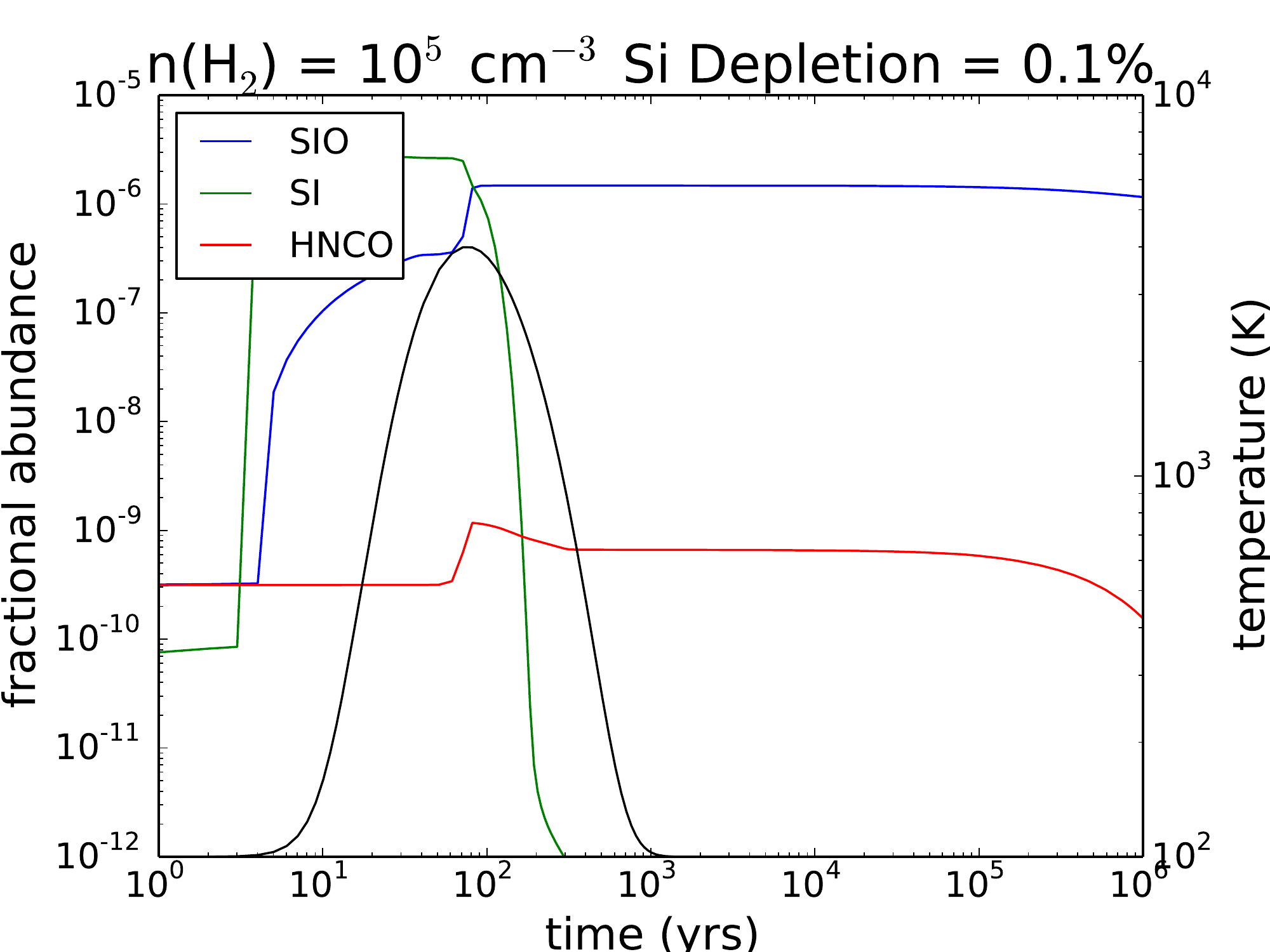}
\includegraphics[width=0.3\linewidth, angle=0]{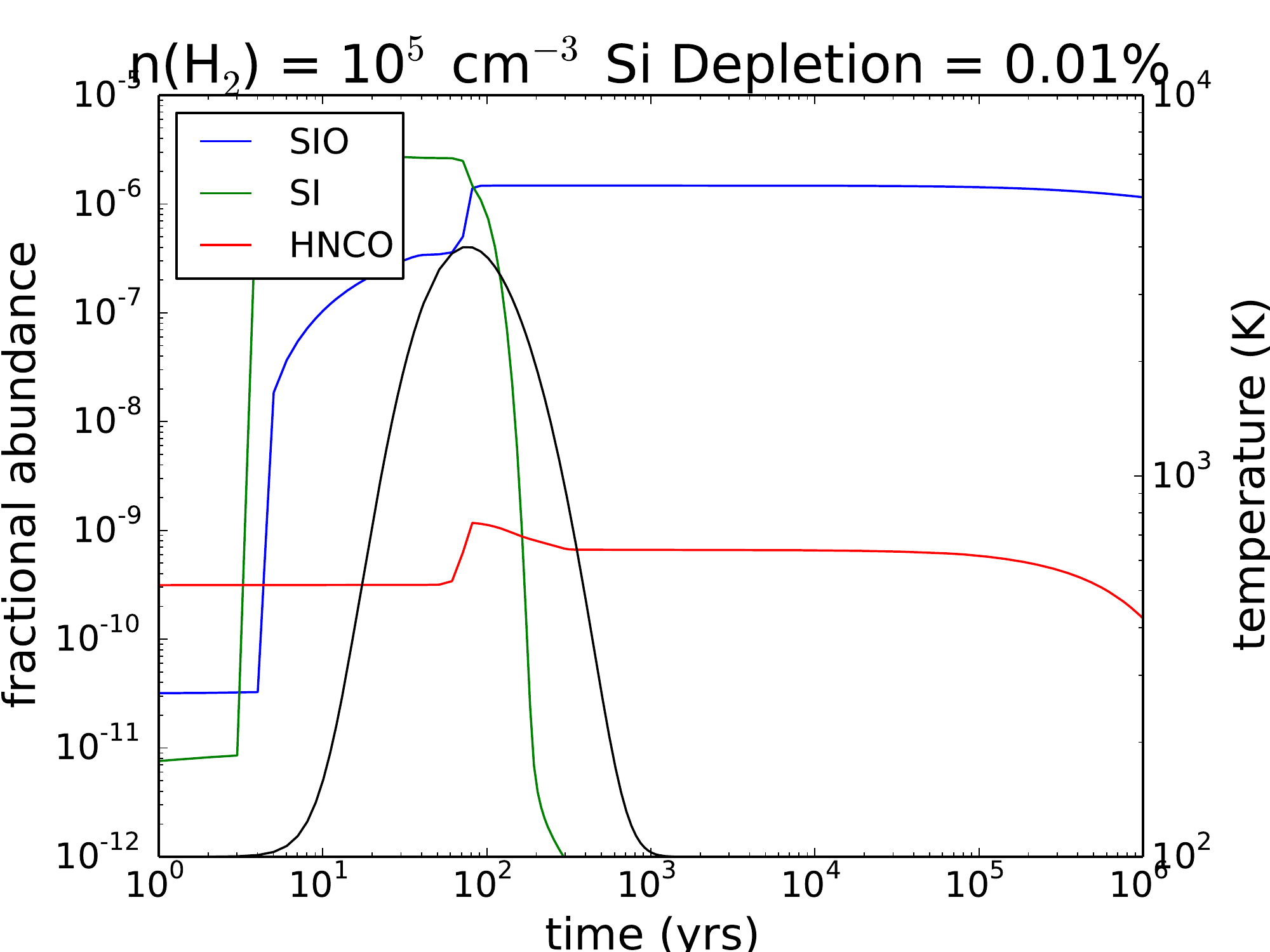}
\includegraphics[width=0.3\linewidth, angle=0]{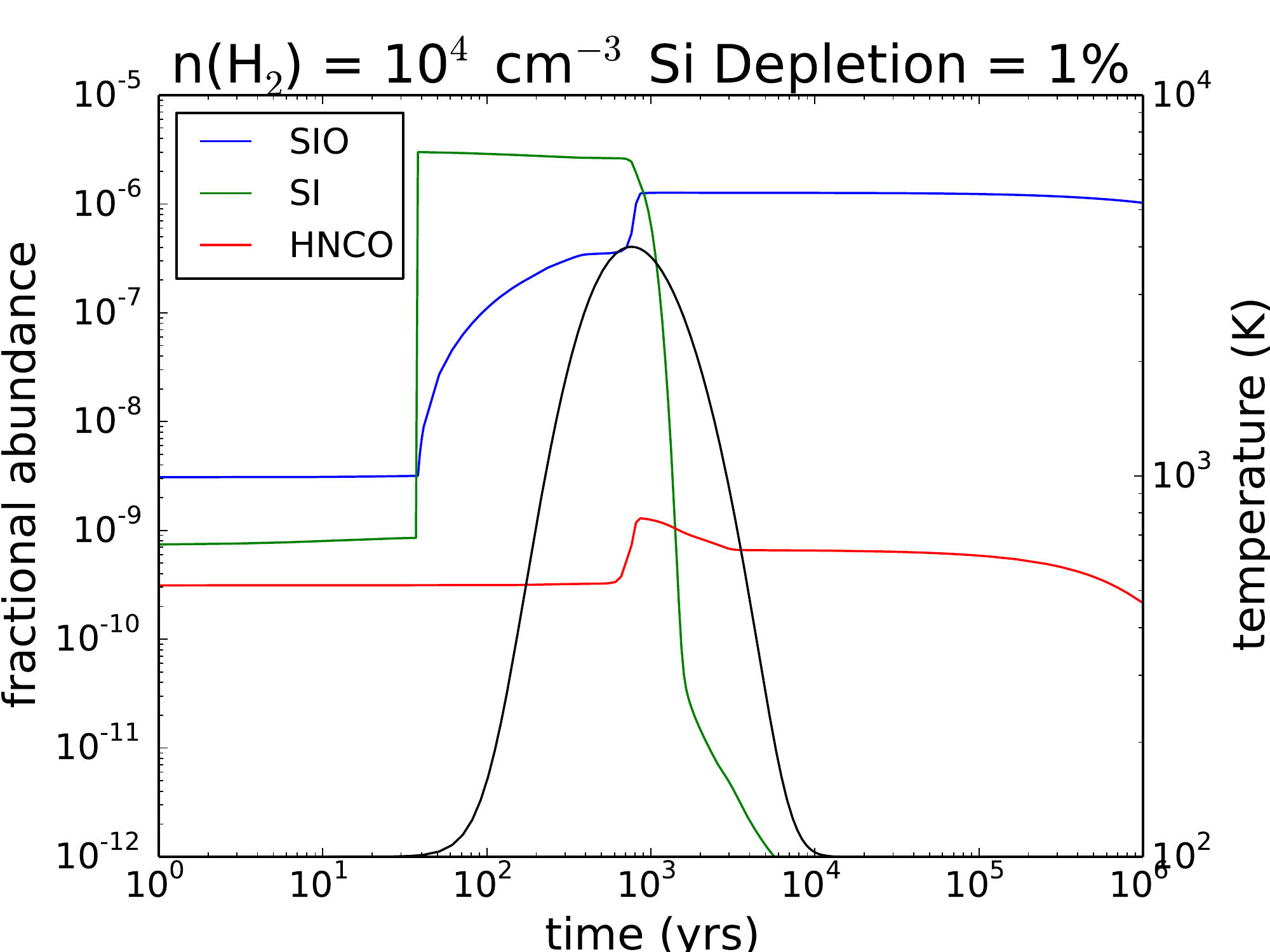}
\includegraphics[width=0.3\linewidth, angle=0]{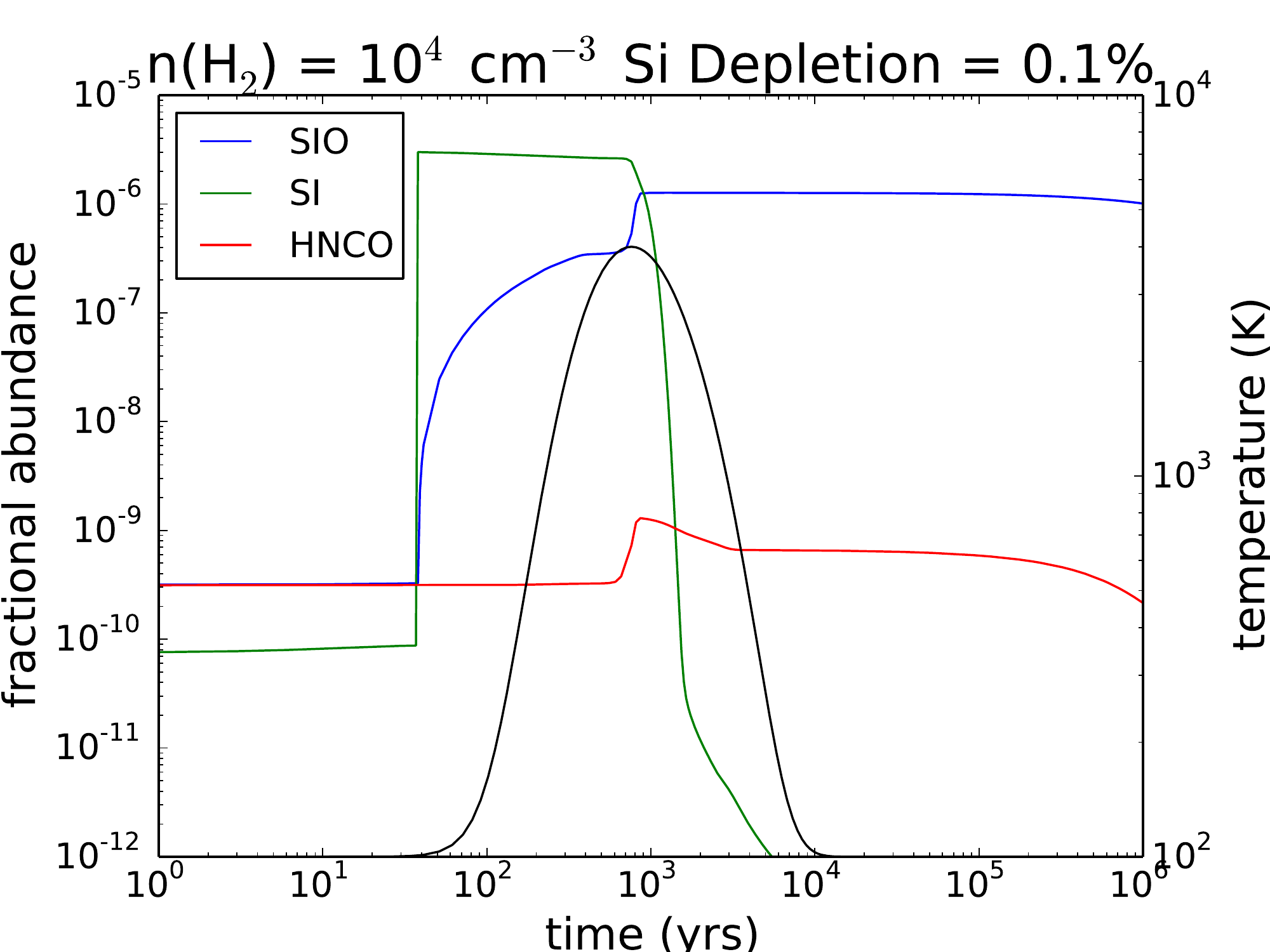}
\includegraphics[width=0.3\linewidth, angle=0]{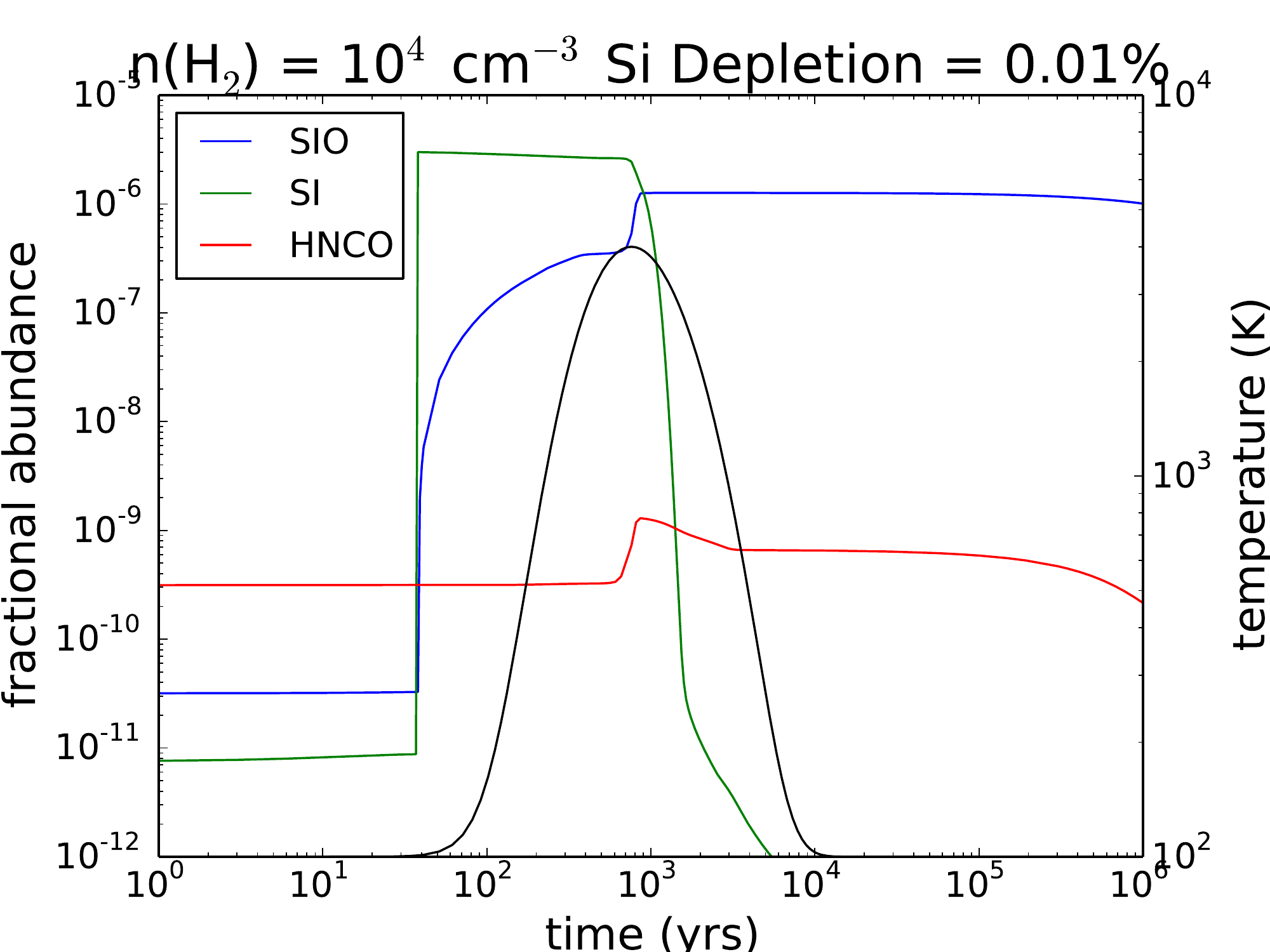}
\includegraphics[width=0.3\linewidth, angle=0]{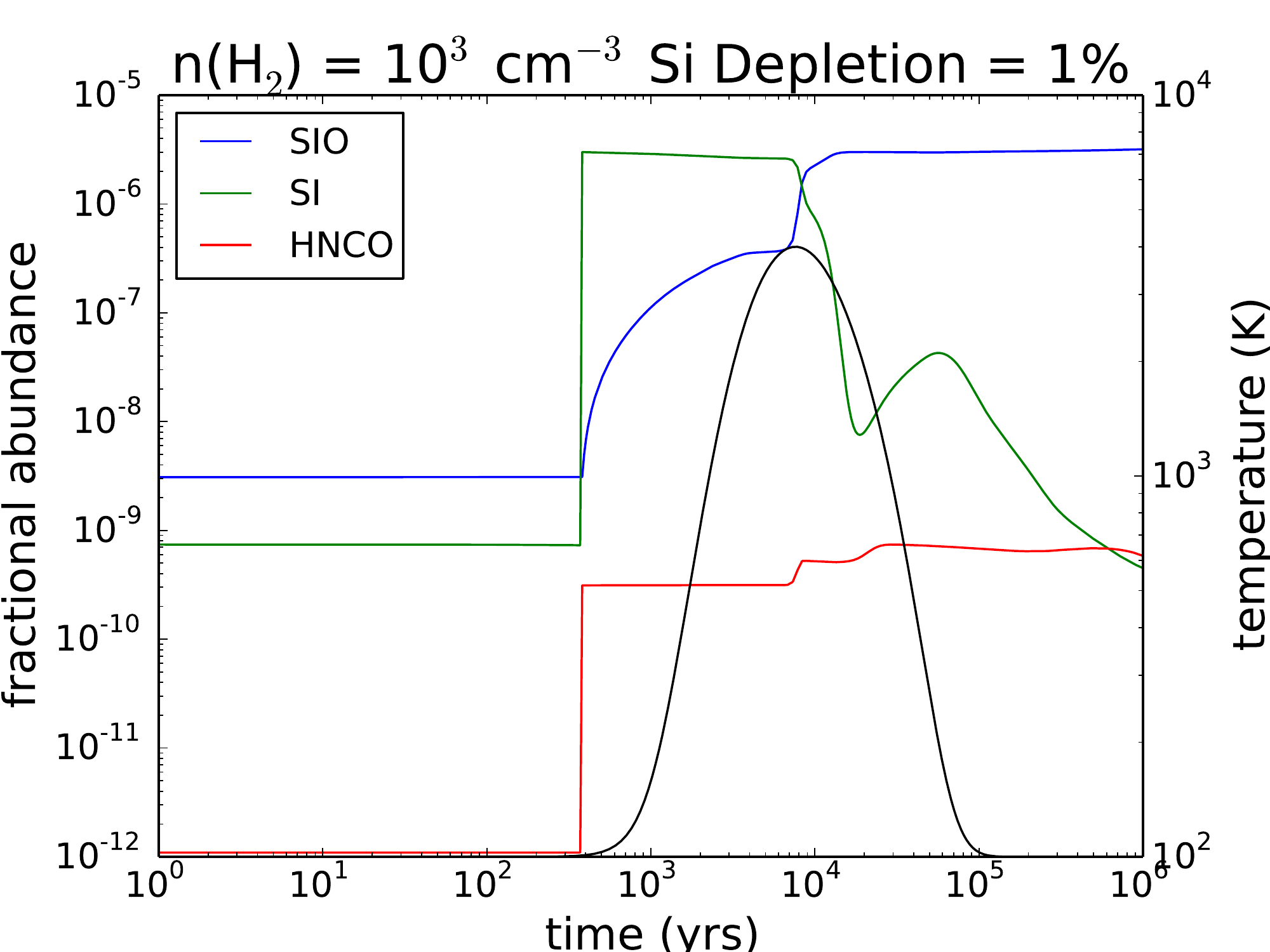}
\includegraphics[width=0.3\linewidth, angle=0]{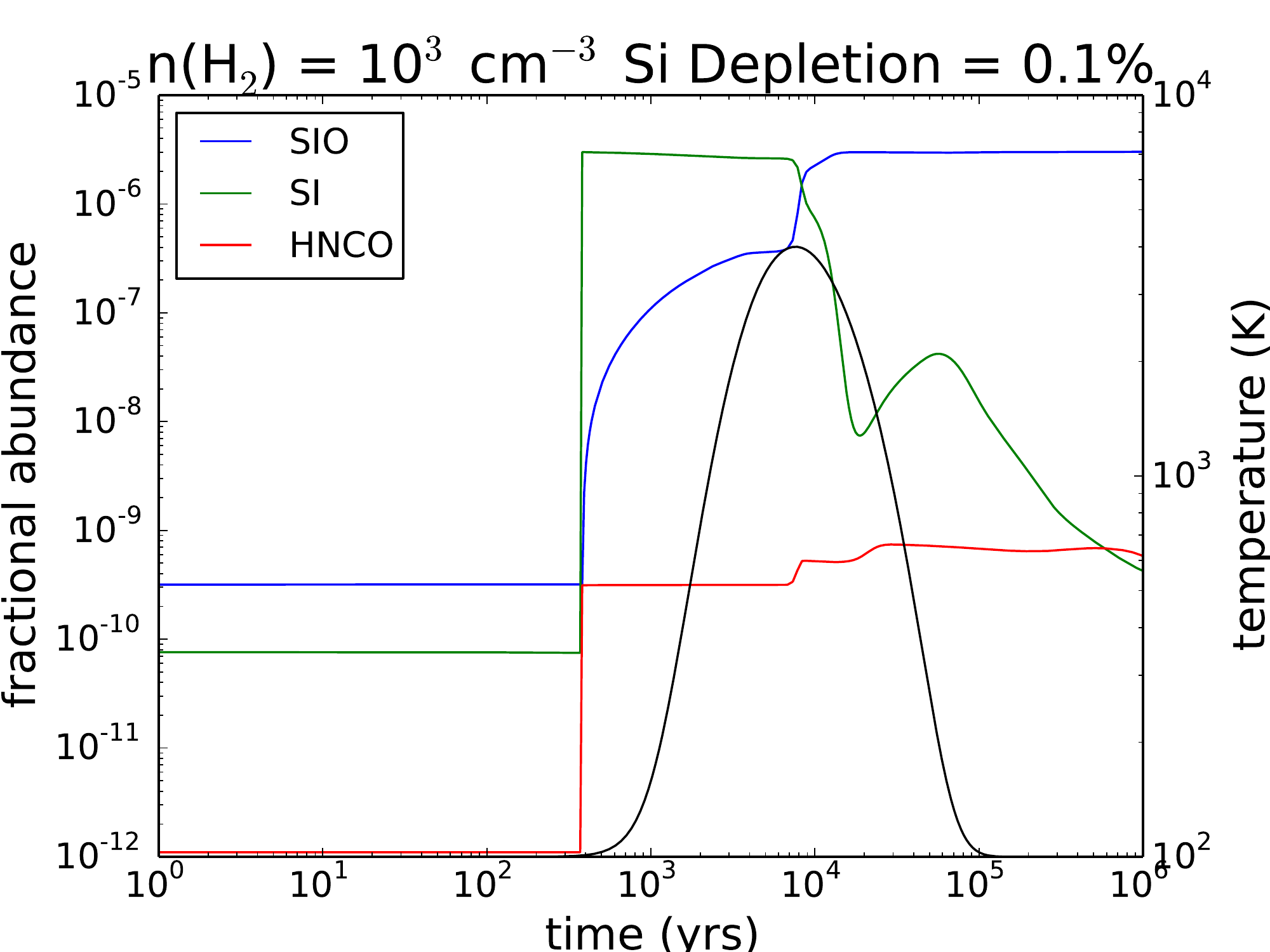}
\includegraphics[width=0.3\linewidth, angle=0]{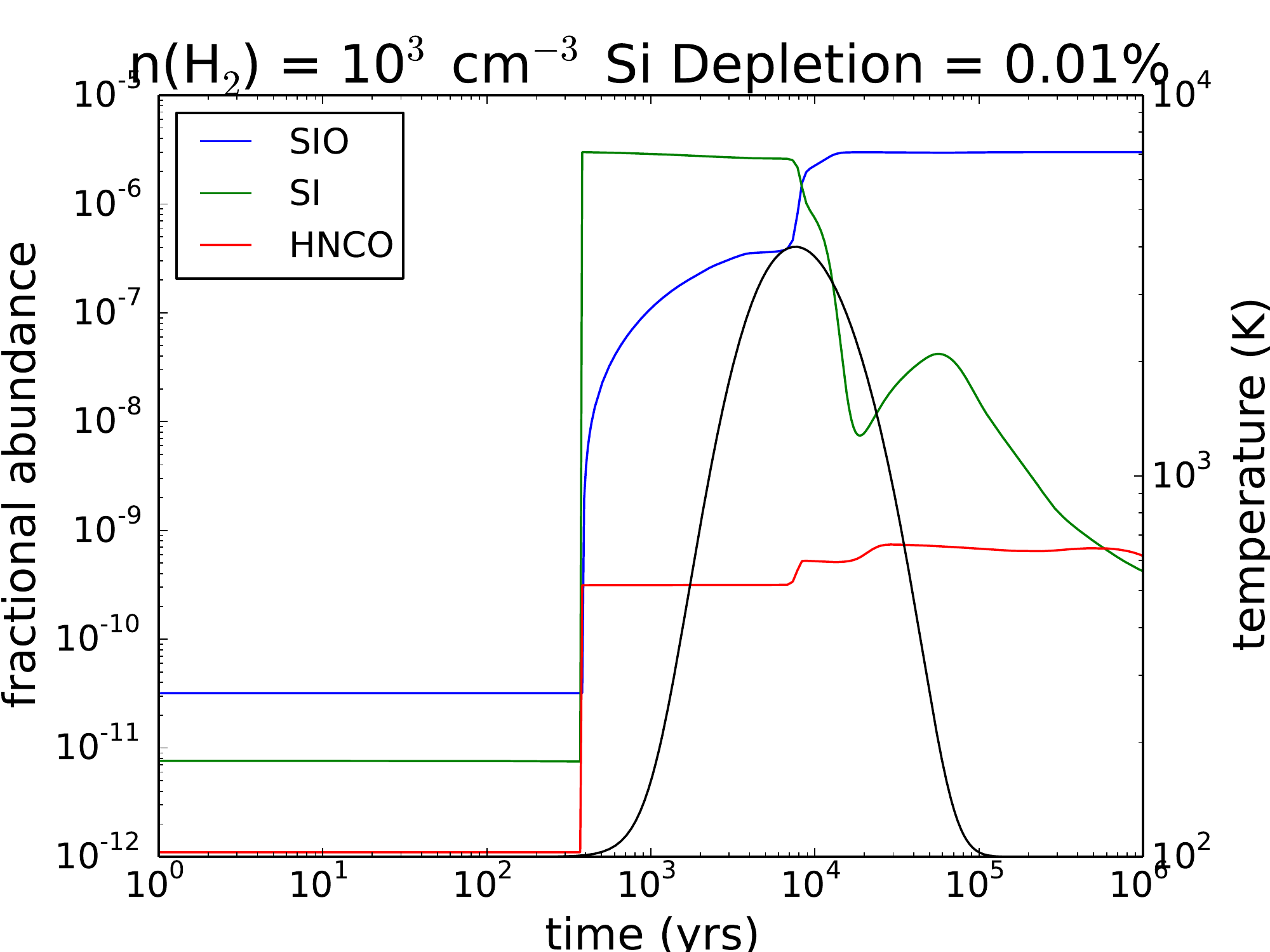}
\caption{Chemical shock modelling showing Si, SiO and HNCO. These are fast (60 km/s) shock models. Bottom: models 1-3. Middle: models 4-6. Top: models 7-9. The black line shows temperature variation.} 
\label{fig:chemmod1}
\end{figure*}

\begin{figure*}[htbp]
\centering
\includegraphics[width=0.3\linewidth, angle=0]{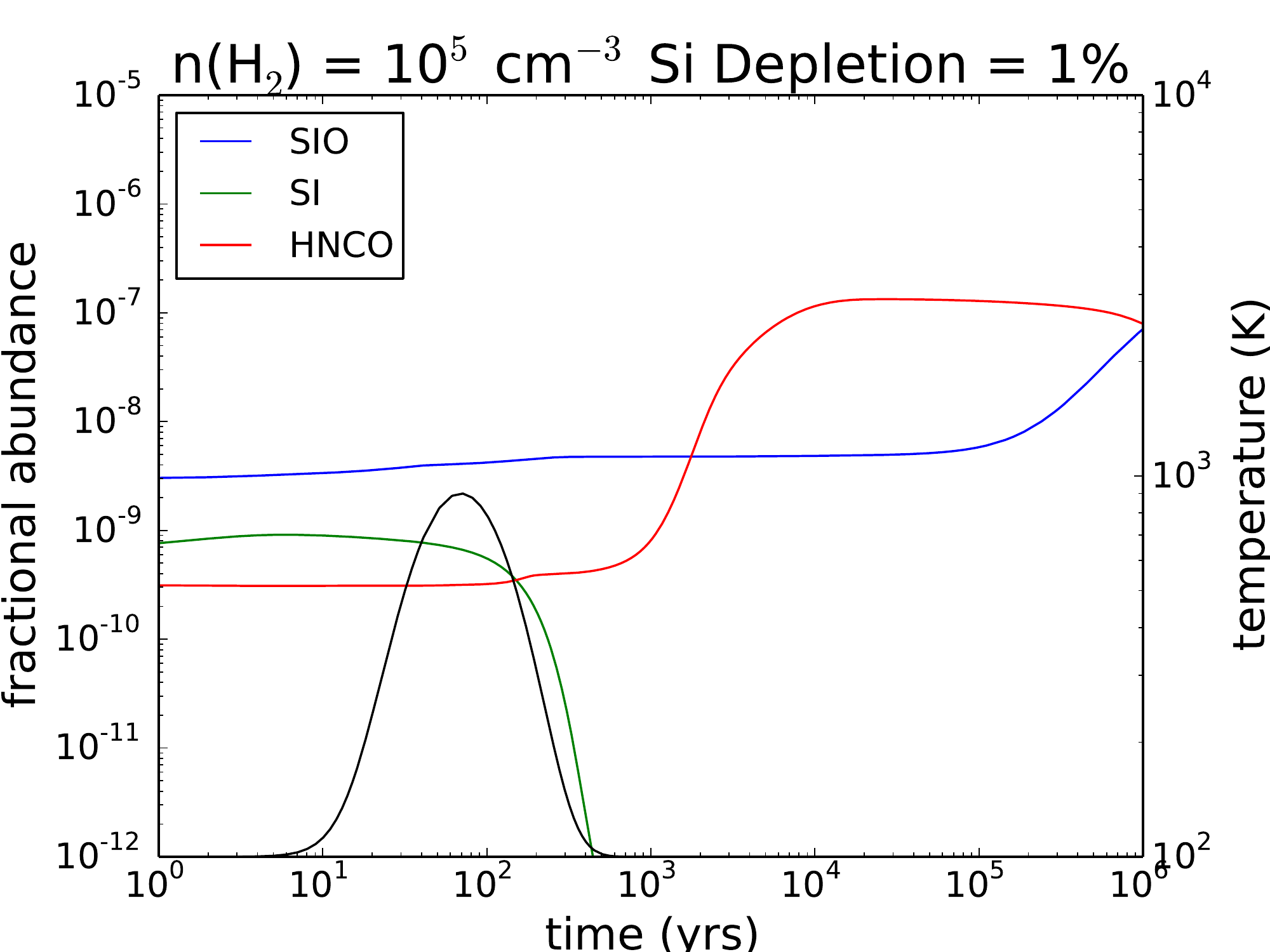}
\includegraphics[width=0.3\linewidth, angle=0]{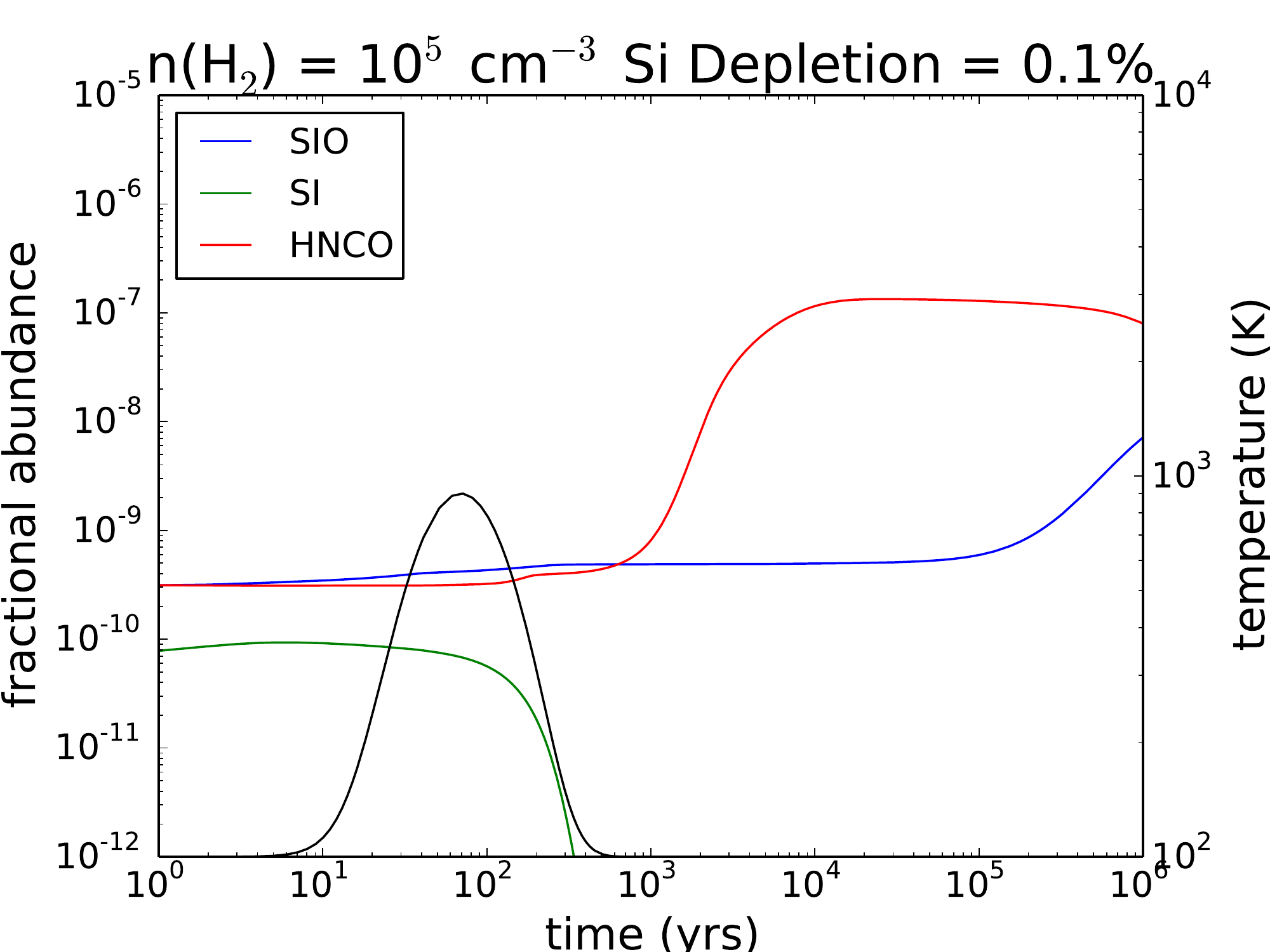}
\includegraphics[width=0.3\linewidth, angle=0]{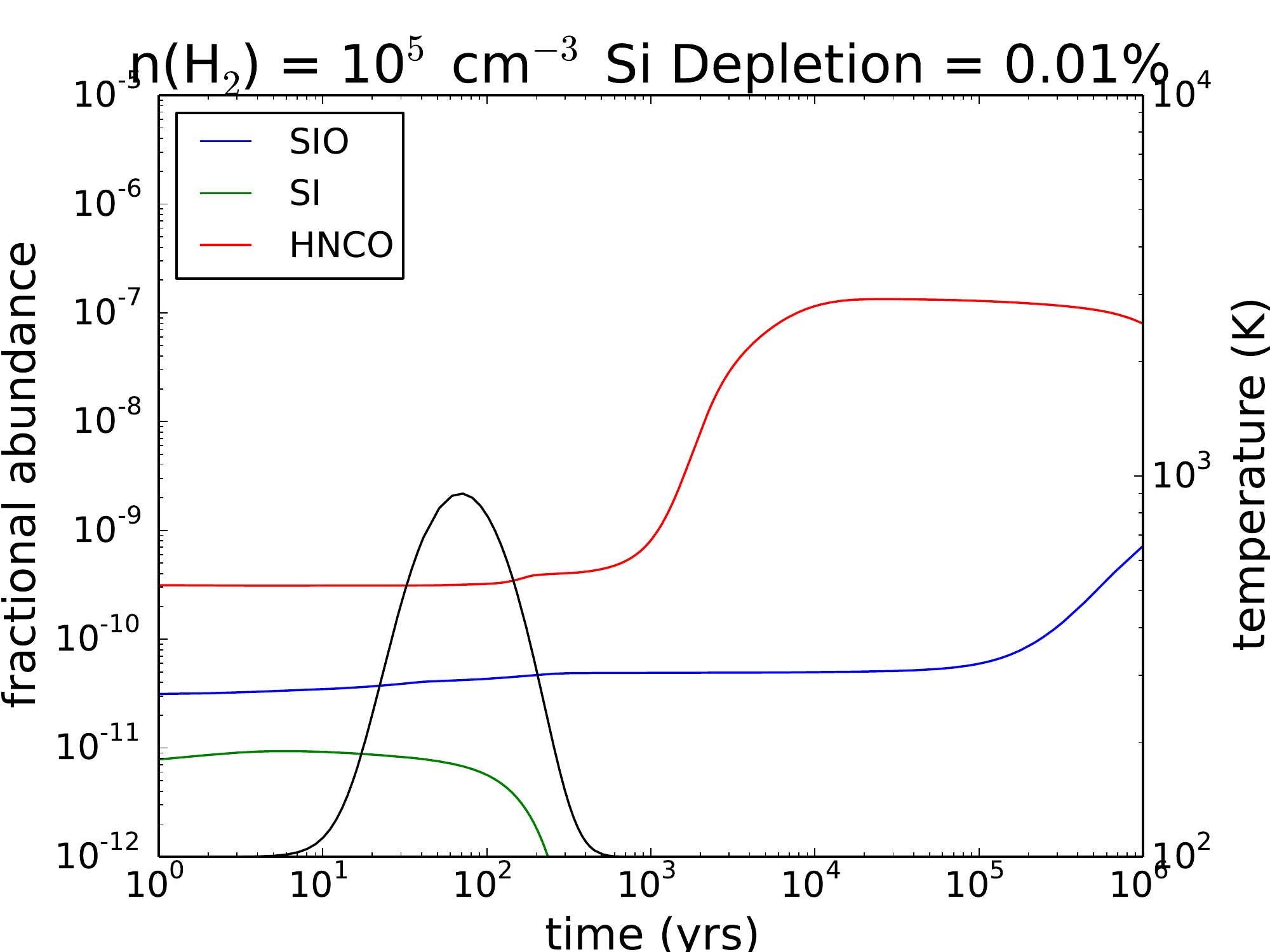}
\includegraphics[width=0.3\linewidth, angle=0]{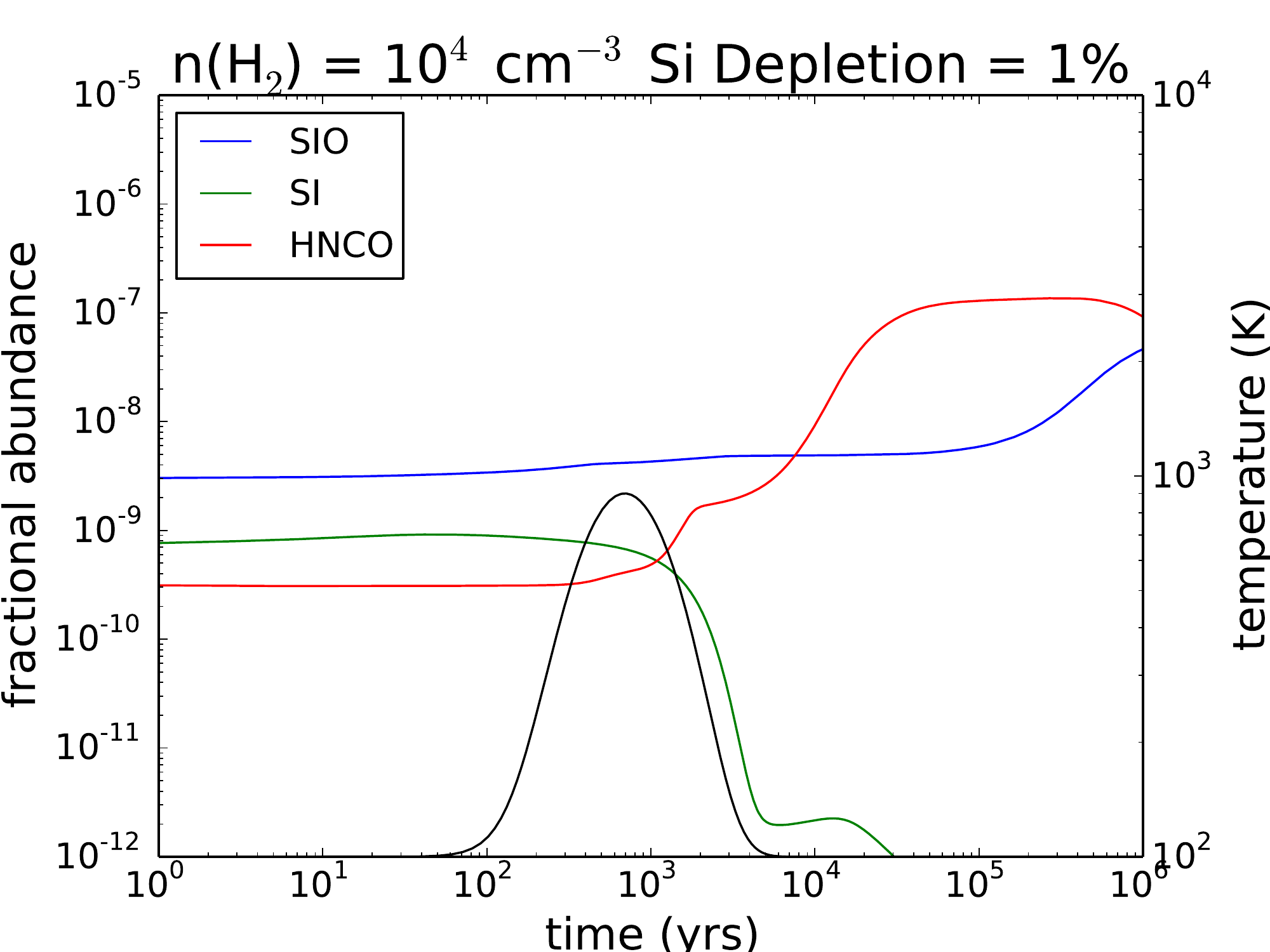}
\includegraphics[width=0.3\linewidth, angle=0]{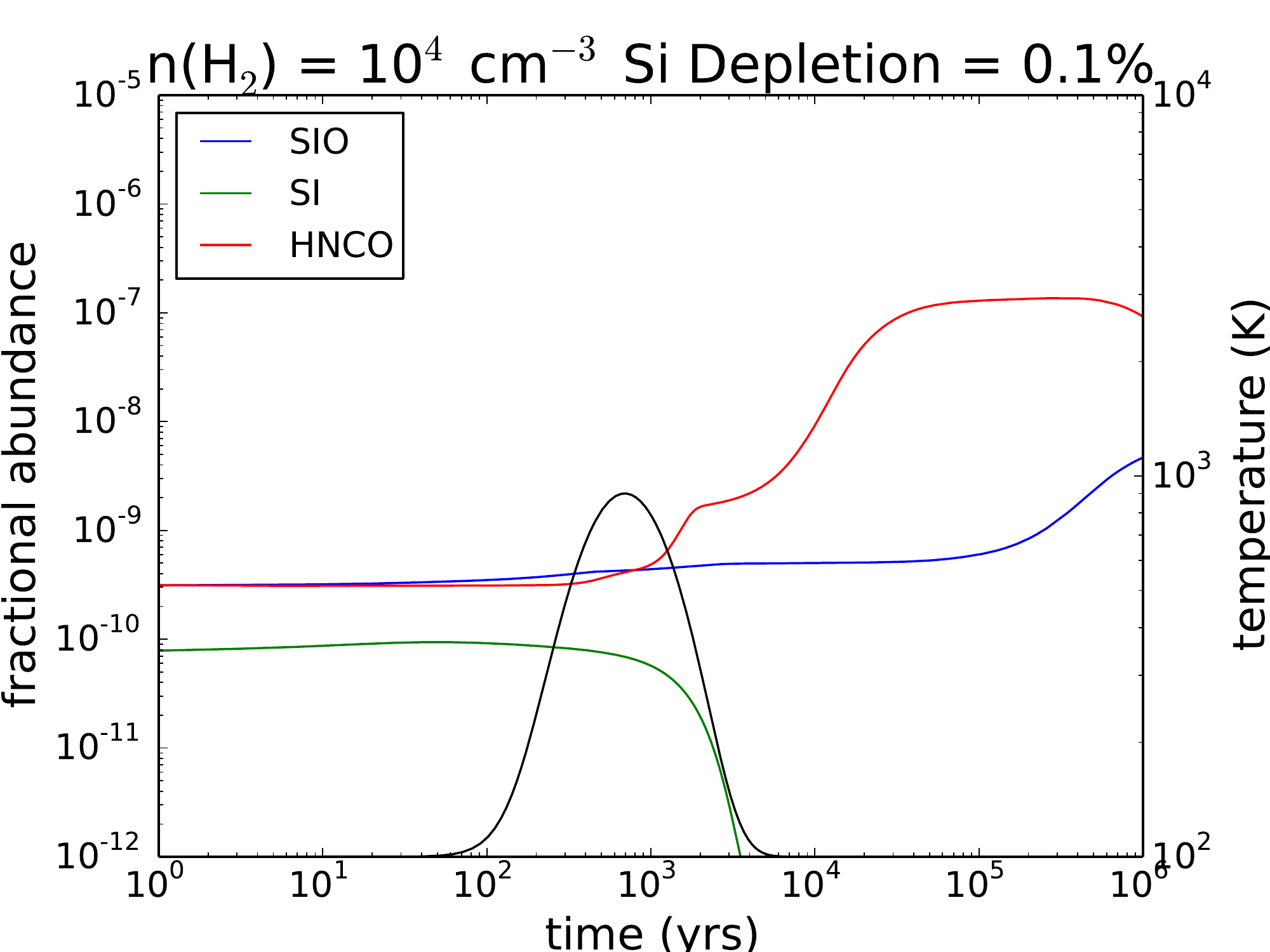}
\includegraphics[width=0.3\linewidth, angle=0]{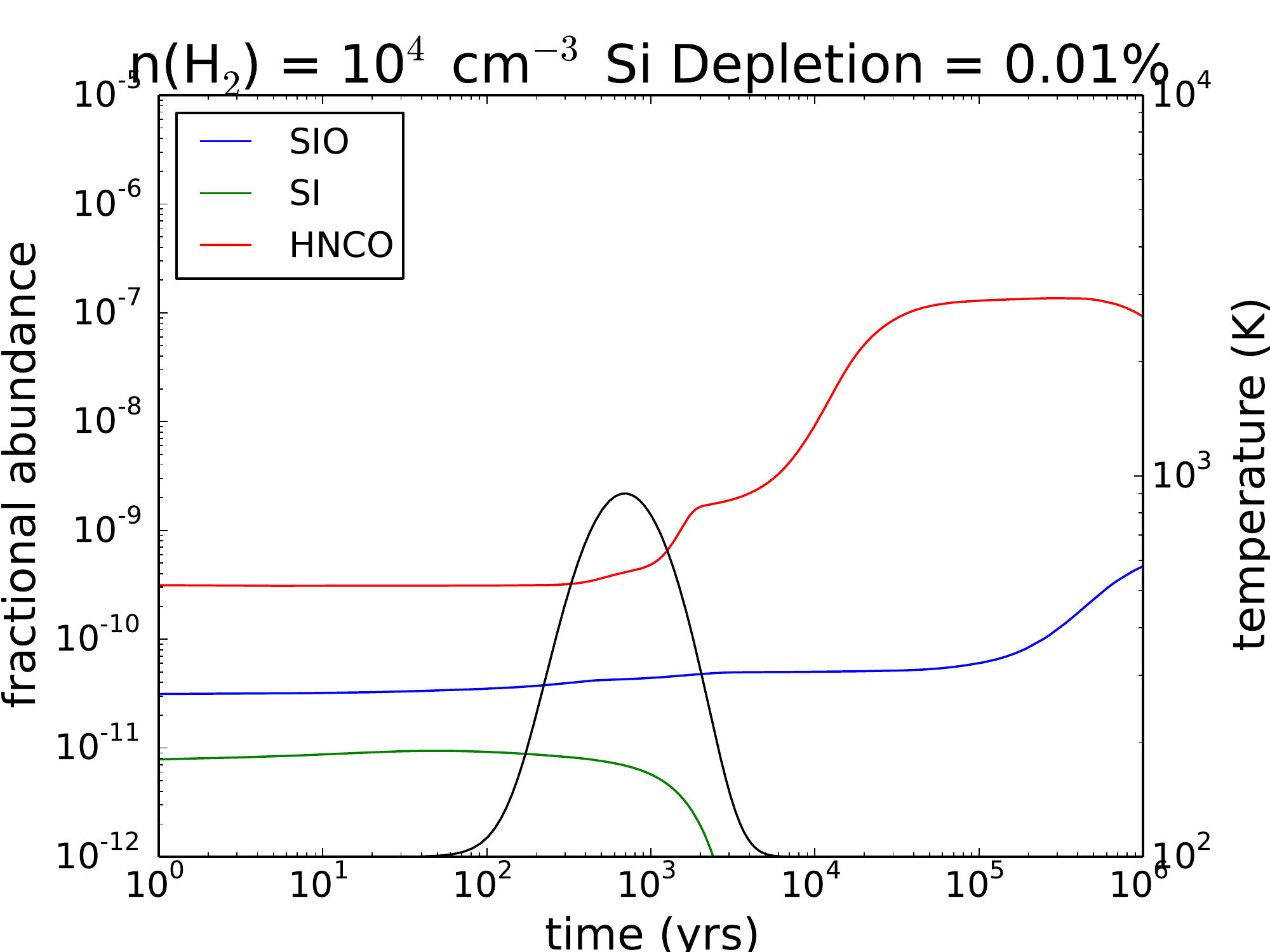}
\includegraphics[width=0.3\linewidth, angle=0]{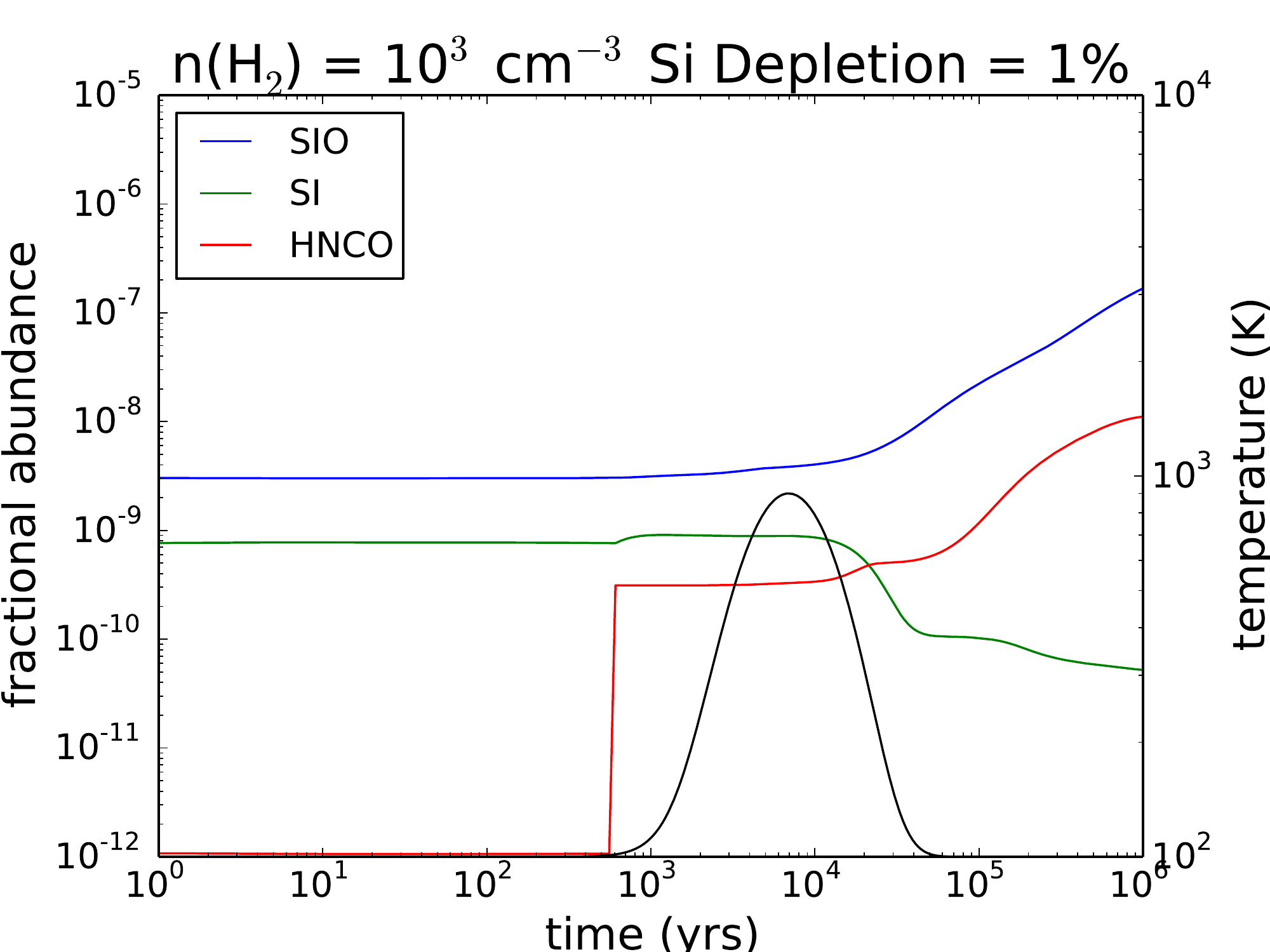}
\includegraphics[width=0.3\linewidth, angle=0]{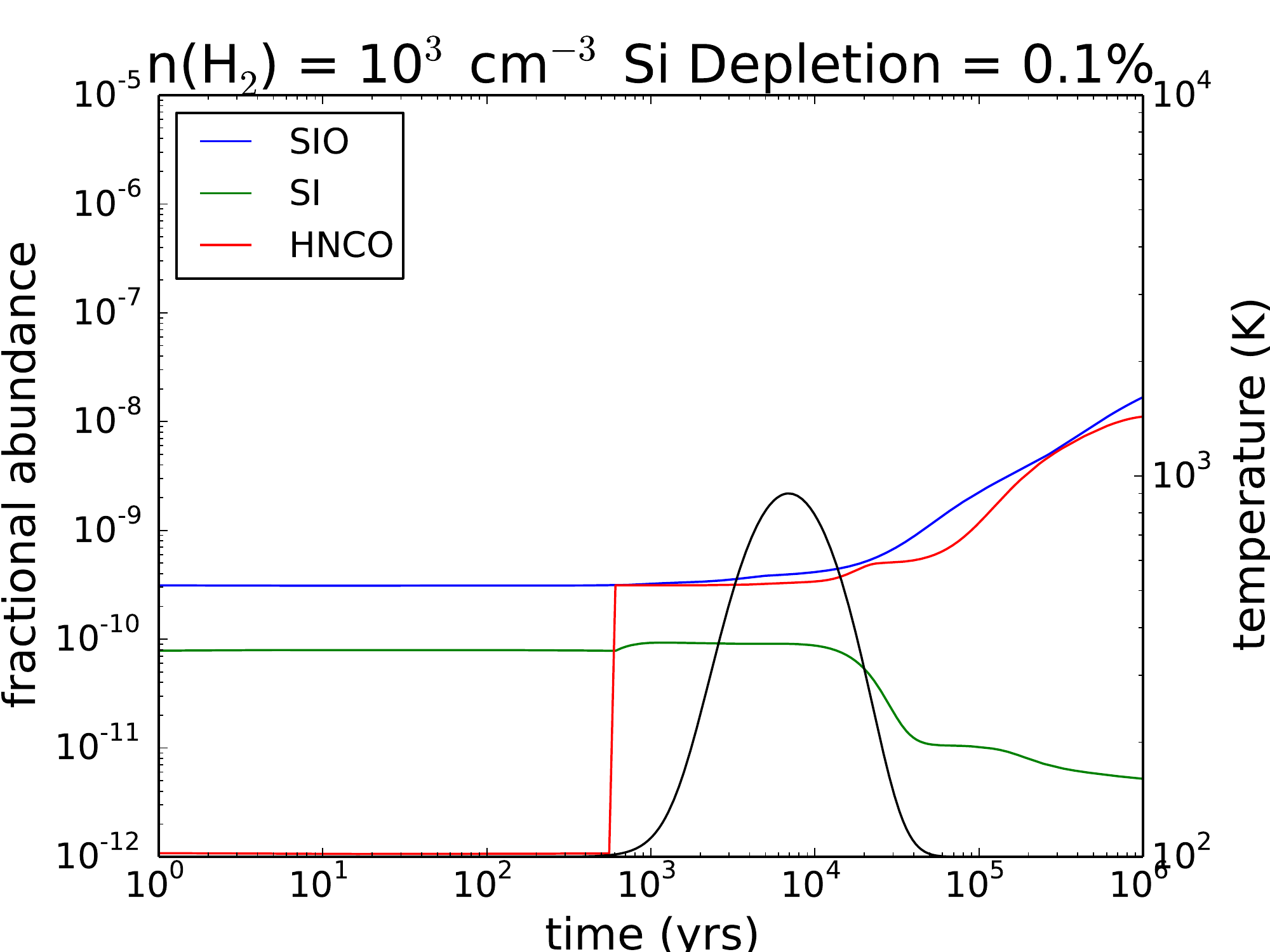}
\includegraphics[width=0.3\linewidth, angle=0]{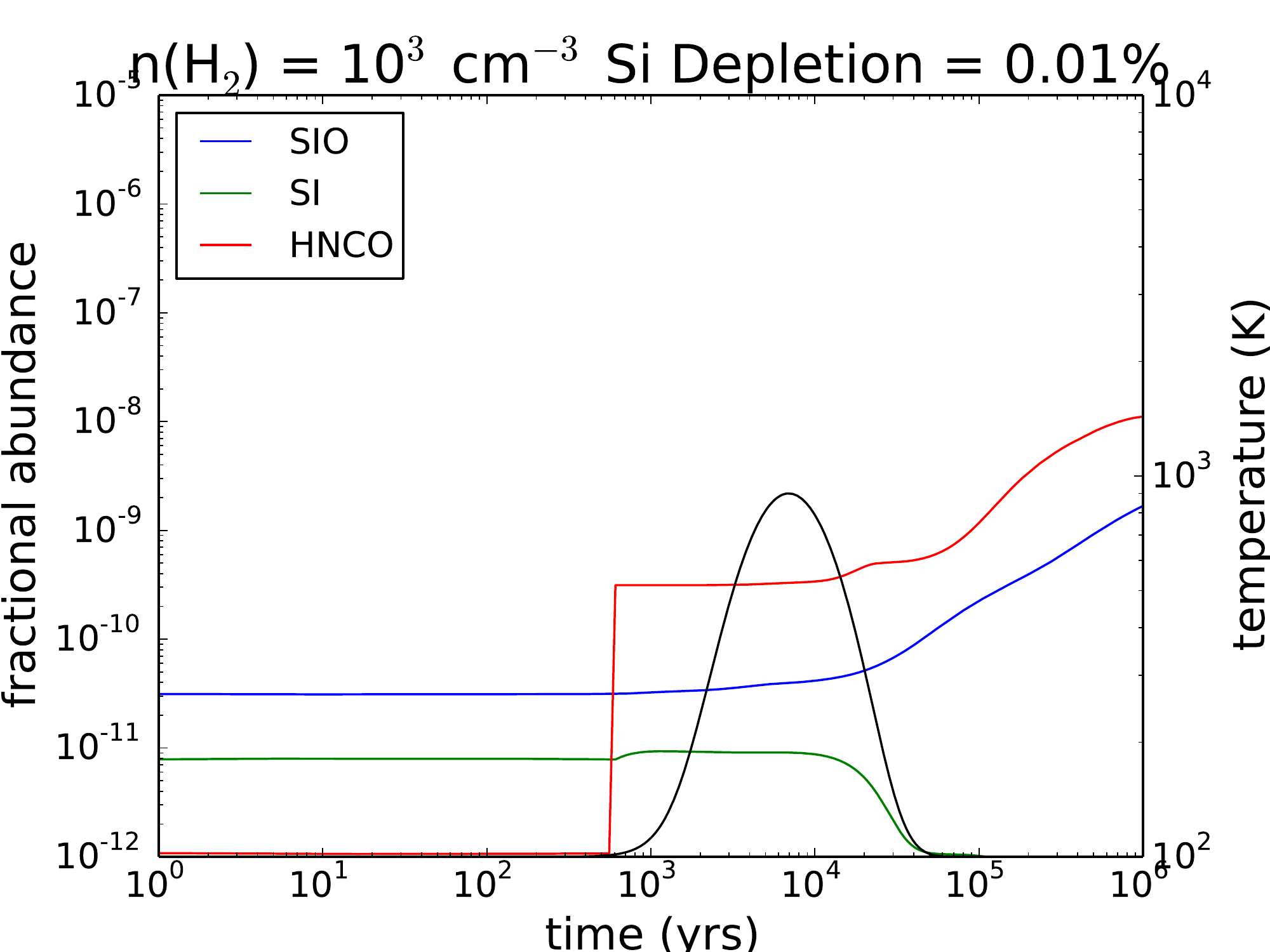}
\caption{Chemical shock modelling showing Si, SiO and HNCO. These are slow (20 km/s) shock models. Bottom: models 10-12. Middle: models 13-15. Top: models 16-18. The black line shows temperature variation.} 
\label{fig:chemmod2}
\end{figure*}

\begin{figure*}[htbp]
\centering
\includegraphics[width=0.3\linewidth, angle=0]{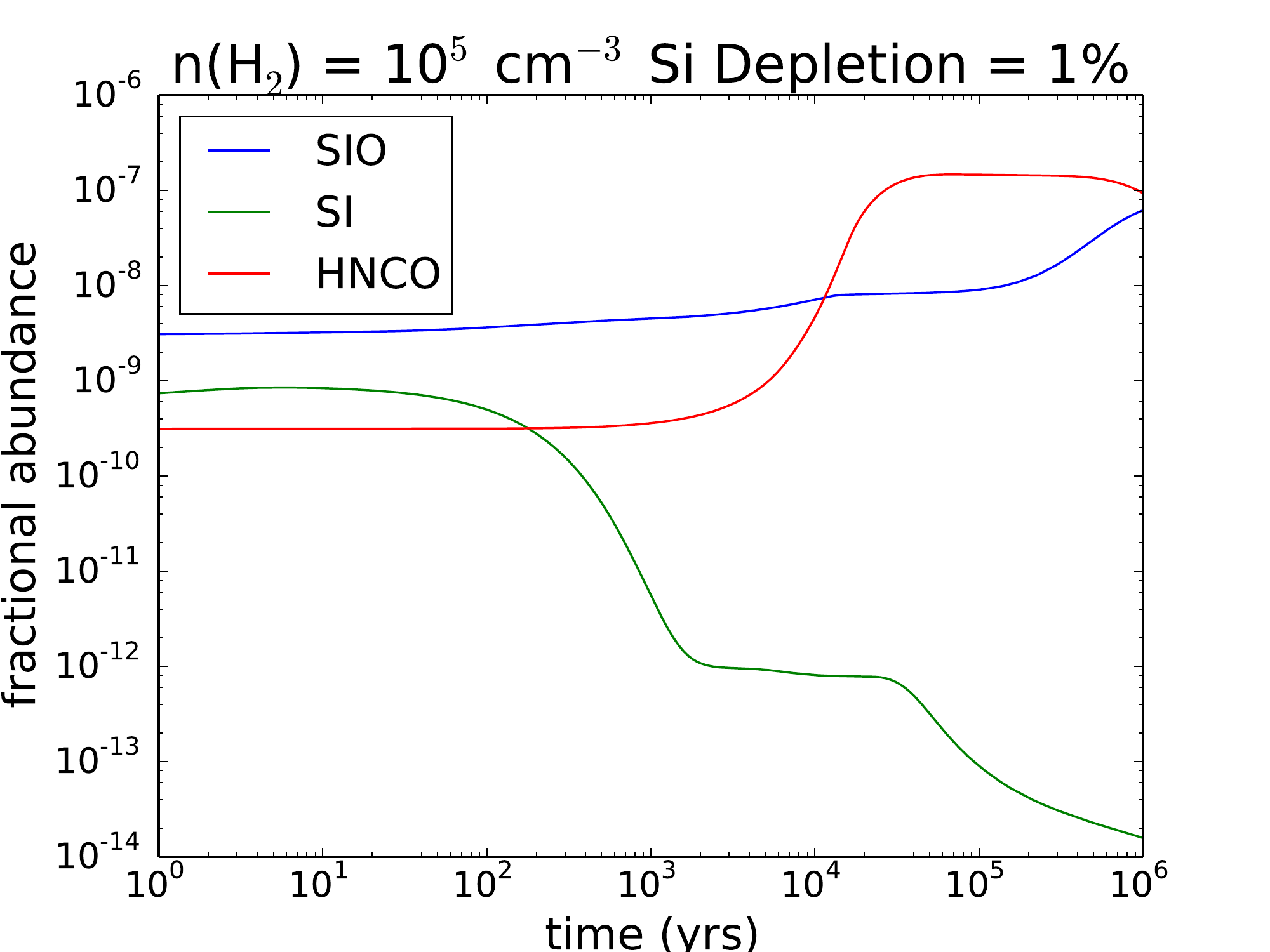}
\includegraphics[width=0.3\linewidth, angle=0]{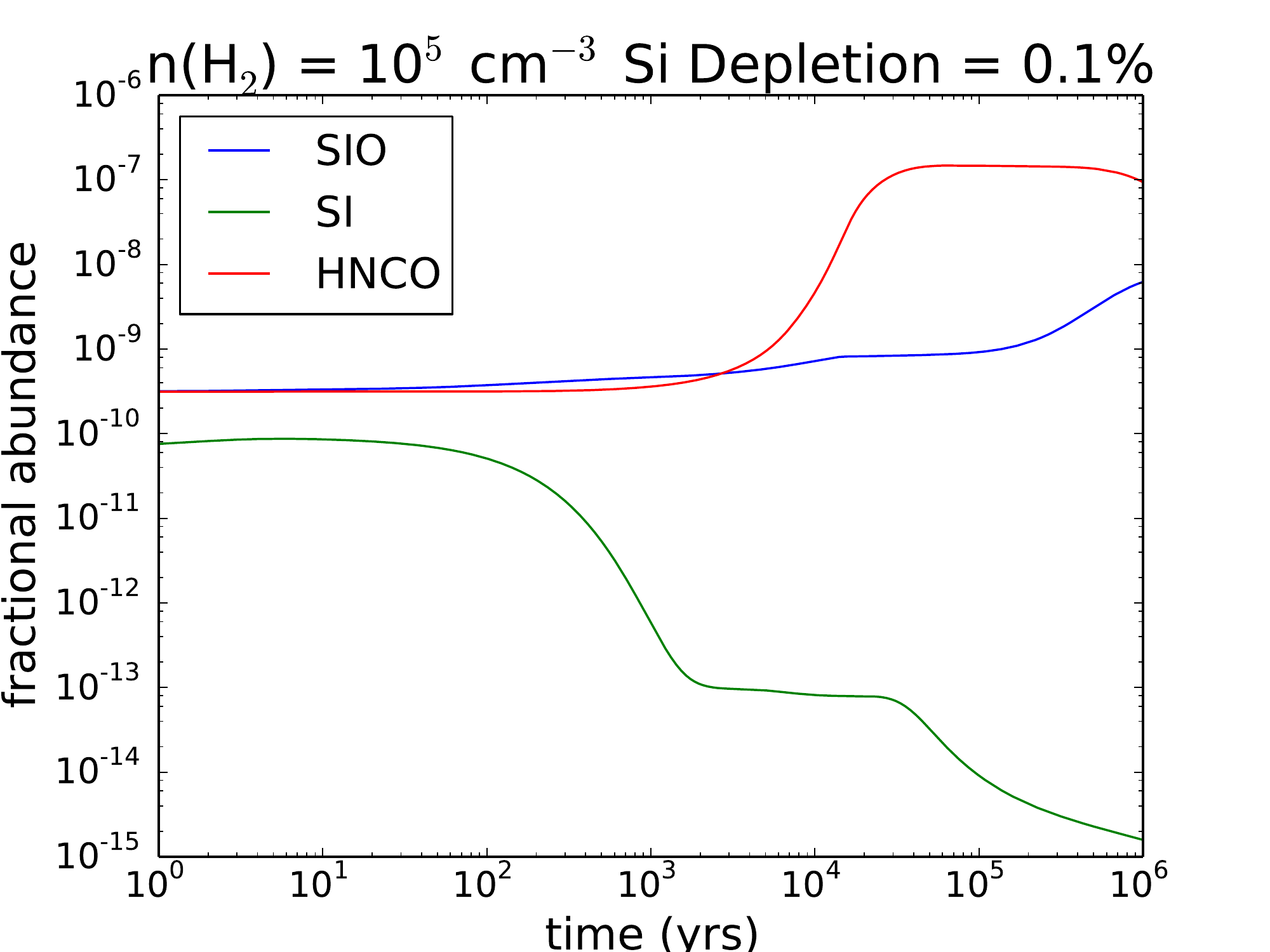}
\includegraphics[width=0.3\linewidth, angle=0]{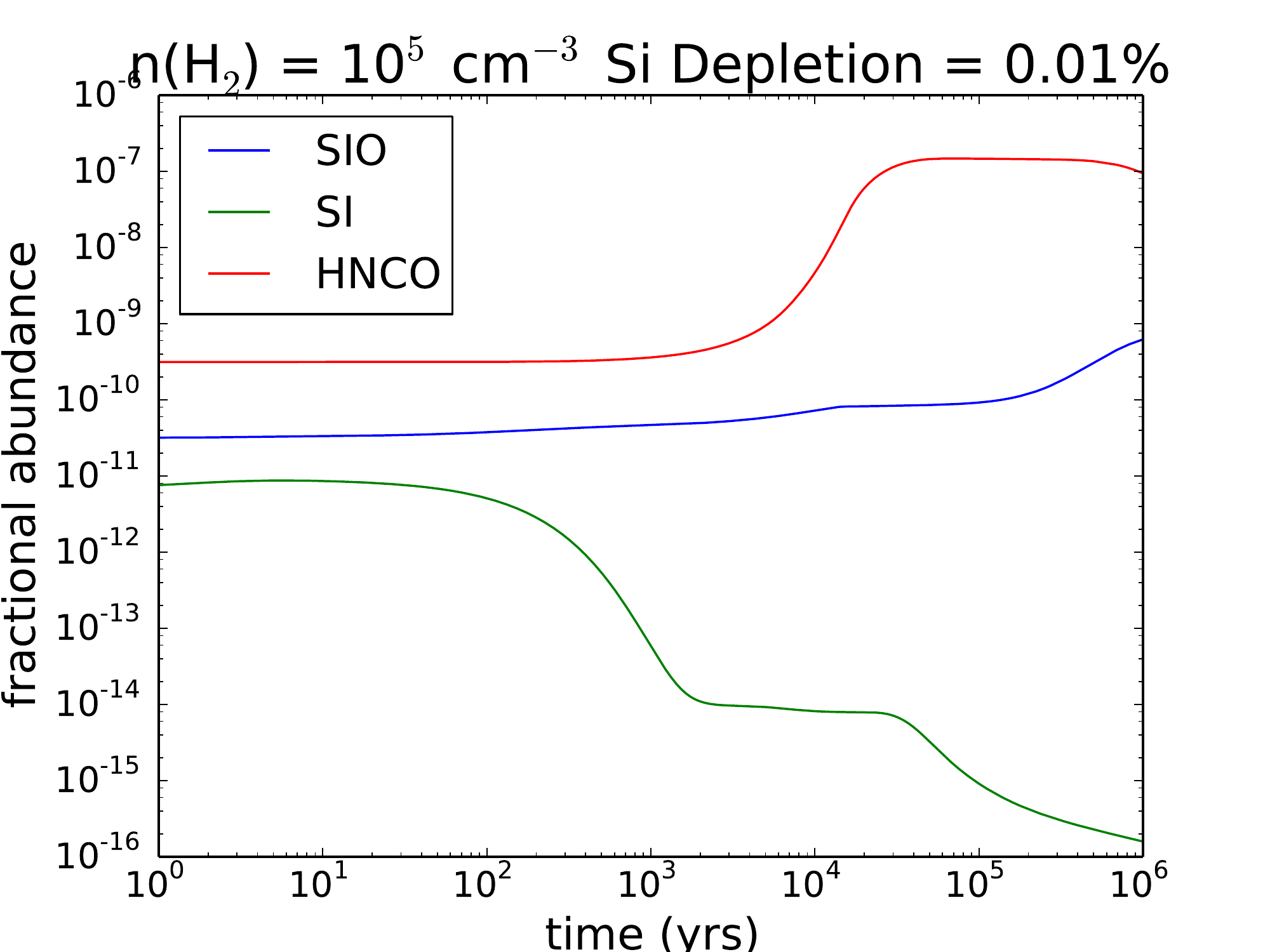}
\includegraphics[width=0.3\linewidth, angle=0]{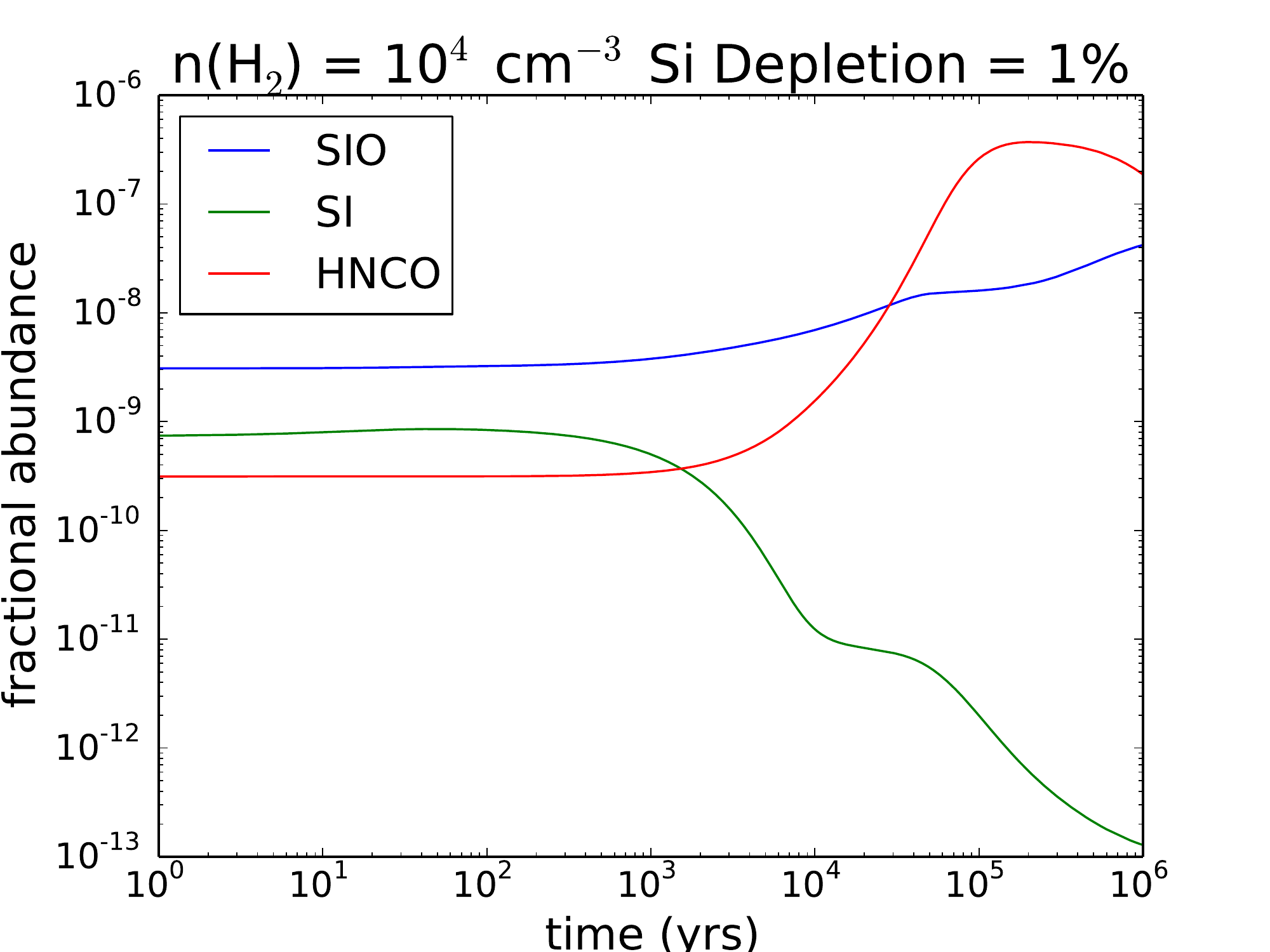}
\includegraphics[width=0.3\linewidth, angle=0]{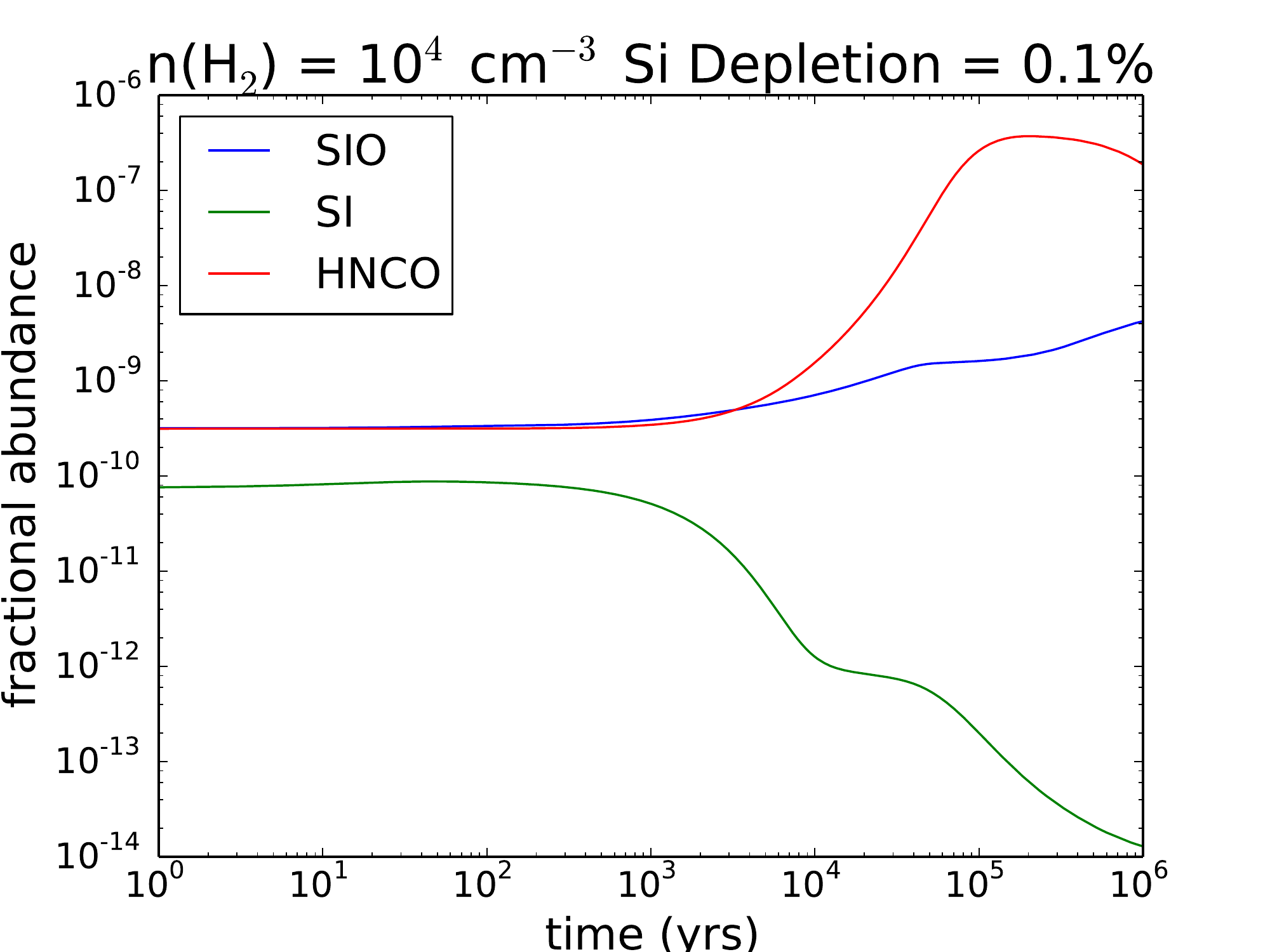}
\includegraphics[width=0.3\linewidth, angle=0]{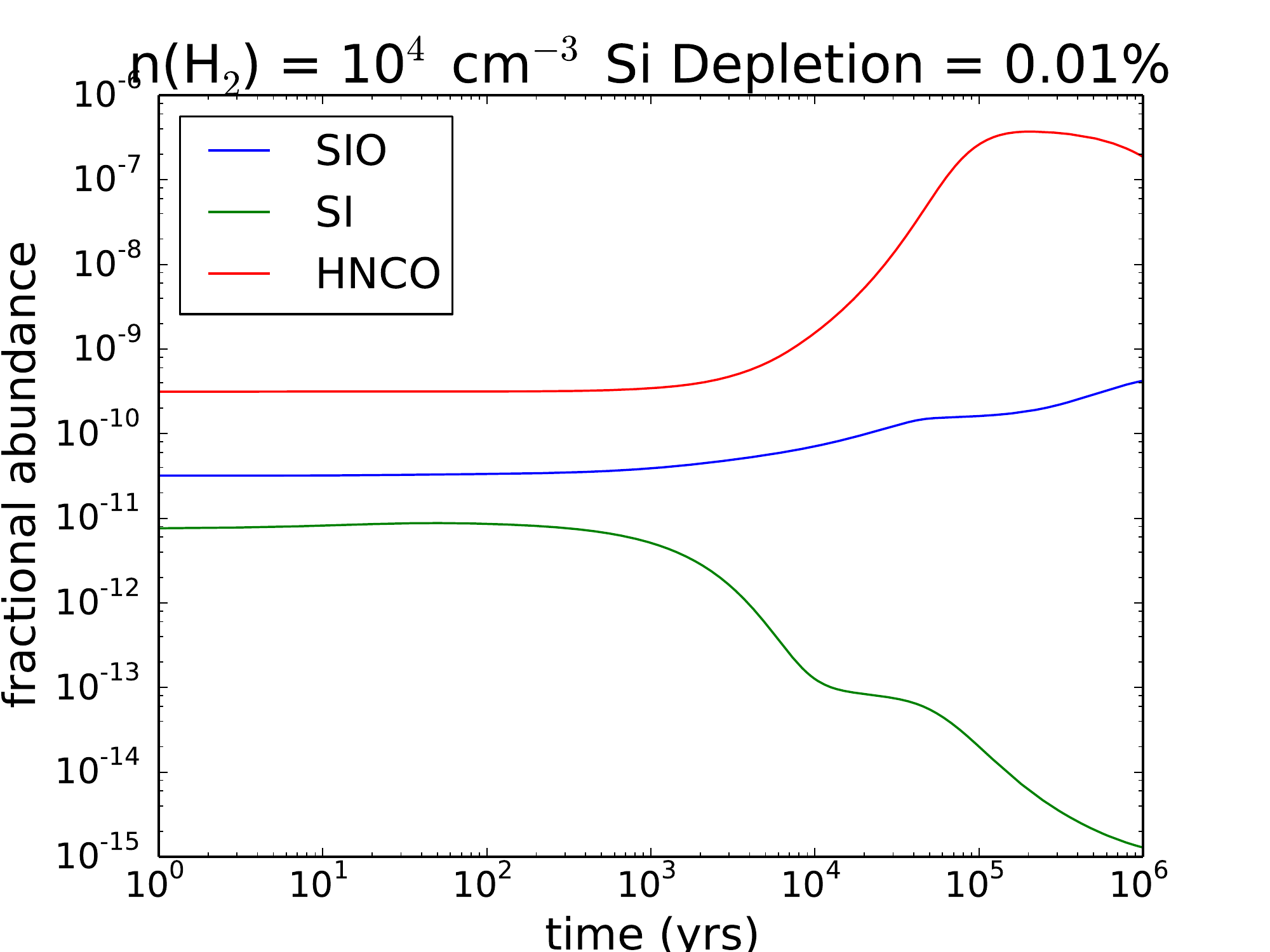}
\caption{Chemical modelling with no shocks showing Si, SiO and HNCO. Bottom: models 19-21. Top: models: 22-24.} 
\label{fig:chemmod3}
\end{figure*}

We run a grid of models, listed in Tables ~\ref{tab:s-models} and ~\ref{tab:ns-models}. \citet{2014A&A...567A.125G} find that the outflow in NGC 1068 has velocities up to 100 km s$^{-1}$, but which vary strongly over the regions observed. The area is large and we are not just observing a single shocked area. As a representative fast shock, we take a value of 60 km s$^{-1}$ for our models. At this velocity, some grain sputtering occurs; the percentage of Si from the grains that gets sputtered is debated: we use the results from the modelling of \citet{2008A&A...482..549J}, see Figure 6, for the purpose of this study and we shall therefore assume that 10\% of the Si locked in the dust grains will sputter. At these velocities, all the mantle is sputtered.

We also investigate the possibility that there are some much weaker shock events occurring, and take a value of 20 km s$^{-1}$. At this velocity, sputtering from grain nuclei is negligible but mantles should return to the gas-phase quite efficiently \citep{2008A&A...490..695G}. In addition, we run non-shock models, as a check to see if we could produce observable quantities of both species without the need for shock chemistry. Note that the non shock models are still run at a temperature (typical of regions where stars have formed) that lead to mantle sublimation. Most silicon is locked up in grain nuclei and silicon is therefore significantly depleted from solar values. As much as 99.96\% of silicon may be depleted into dust grain cores \citep{2008A&A...482..549J, 1996ApJ...468L..65S, 1989GeCoA..53..197A}. This value has some uncertainly so we adopt three values for our initial abundance of Si of X(Si) = 1\%, 0.1\% and 0.01\% of the solar value. {We hold the value of the UV radiation field at the standard interstellar value. The high extinction in the dense gas we model, even pre-shock, would shield the gas so that an increase in UV would have little effect \citep{2014A&A...570A..28V}.}

Krips et al. (2011) as well as Viti et al. (2014) find that the gas in the regions we are observing has a density, $n$(H$_{2}$) $\ge$ \num{e4} cm$^{-3}$ (also confirmed by our RADEX analysis) and that the average gas temperature between 60 and $\sim$ 200 K. We choose to run our shock models with a pre-shock density of \num{e3}-\num{e5} cm$^{-3}$ (leading to a post-shock density of $\ge$ \num{e4} cm$^{-3}$). At \num{e3} cm$^{-3}$ the gas and the dust is not coupled. Therefore the temperature of the dust will be lower than the gas temperature, and at this density the dust grain mantles are not sublimated until the shock occurs and temperature and density increase. At higher densities, we assume coupling between gas and dust and hence the mantles are already sublimated before the shock occurs. Our non shock models are completed using density \num{e4}-\num{e5} cm$^{-3}$.

In Figure ~\ref{fig:chemmod1} we show the fractional abundance (with respect to the total number of hydrogen nuclei) of Si, SiO and HNCO for our fast shock models. The main difference between the models with a very low pre-shock density and the higher densities ones are that for the former there is no mantle sublimation until the shock arrives, while in the latter the mantle thermally sublimates; this leads to a slightly different behaviour for HNCO which always undergoes an enhancement during the shock passage: however for higher pre-shock densities it also seems to decrease again once the shock has passed.  In all the fast shock models Si is significantly enhanced as the shock sputters the grain core. The Si, now in the gas phase, reacts quickly with O$_2$ {and OH} during the shock event. {In addition, at the peak of the shock, Si reacts with CO to form SiO, the very high temperatures overcoming its large activation barrier. These reactions cause} a rapid rise in SiO. SiO abundance levels off after the shock but Si falls, reacting in particular with C$_2$H$_2$ to form SiC$_2$. The amount of Si initially depleted into olivine grain cores does not have a large impact on the final abundance of SiO. This is because much of the Si depleted returns to the gas due to sputtering during the shock.

In Figure ~\ref{fig:chemmod2} we show our weak shock chemical models. For the weak shock models the main difference, due to the different densities used, is on whether the gas and dust are coupled: in the latter case sublimation of the icy mantles occurs before the shock arrives, as in the case of the fast shock models, while for the low density models the icy mantles are released back to the gas phase only when they get sputtered by the passage of the shock. However, because the density is very low (10$^3$ cm$^{-3}$) hardly any freeze out takes place in Phase 1. Sputtering of the grain cores does not occur. In both low and high density models 
we find that following the shock the HNCO abundance increases by up to 3 orders of magnitude. The increase is not so pronounced for the SiO abundance. In this case the amount of Si depleted to the dust is important in determining the final SiO abundance, as the weaker shock does not sputter the grain core.

In Figure ~\ref{fig:chemmod3}, we show our chemical models without any shocks. We see here a very small rise in SiO over time. HNCO increases significantly at late times. These models were ran at a density always equal or higher than 10$^4$ cm$^{-3}$; hence the gas and dust are assumed to be coupled and mantle sublimation therefore always occurs. The initial elemental Si depletion is important in determining final SiO abundance, as nothing returns to the gas phase from the dust grain cores. {We can calculate a fractional abundance from the column densities we calculated from our RADEX modelling to compare to our chemical models. This is subject to two important limitations. Firstly, we must assume a CO/H$_{2}$ fraction. We take the canonical value of \num{e-4}, which also matches well with what we find in our models. In addition, the chemical modelling only gives fractional abundances at single points in time and space. Our observations encompass a very large region, which will include a large gradient of abundances. So while the chemical modelling is very useful for seeing trends in production or destruction of different molecules, it is difficult to quantitively and directly compare to observations. Nevertheless, we calculate the fractional abundance X(SiO) $\approx$ 10$^{-8}$ and X(HNCO) $\approx$ 10$^{-9}$ in the East Knot locations. The column density estimates in the West Knot are very broad but give X(SiO), X(HNCO) $\approx$ 10$^{-11}$ to 10$^{-8}$. The values found in the models can best be described as an upper limit given that our observations cover an area greater than a single shocked region. In the fast shocked models, X(SiO) peaks at $\sim$ \num{e-6}: { this value is one order of magnitude higher than observed in the CND of NGC~1068.} X(HNCO) peaks { at} $\sim$ \num{e-9}, { which is close to the observed value for the fractional abundance of HNCO.} As this is an upper limit, it seems unlikely that the HNCO we observe is being produced significantly in fast shocks. In the weak shock and no shock models, we see the opposite. X(HNCO) peaks at $\sim$ \num{e-7} but X(SiO) only reaches $\sim$ \num{e-8}. This indicates that fast shocks are likely to be producing SiO, whereas weak shocks, or warm, dense gas is likely responsible for the HNCO we observe.}


{To further this analysis, we quantify the difference that the passage of a shock makes by comparing} a {\it fast} shock model with a non shock model ran at the same density (e.g. Model 7 and 22): we find that HNCO is in fact much more enhanced in the non shock models implying that although a fraction does come from the mantles (which evaporate in both cases, at least for the high density models) the bulk of its increase happens in the warm gas phase after the ices sublimate. However, surprisingly, this increase does not occur in the shock models where the temperature is even higher. The main route to formation in the gas phase for HNCO is Reaction ~\ref{eq:HNCO}.

\begin{equation}
\ce{CH_2 + NO -> HNCO + H}
\label{eq:HNCO}
\end{equation}

During the fast shock, NO is very rapidly destroyed through reaction with atomic hydrogen (Reaction ~\ref{eq:NO}), which is enhanced. 

\begin{equation}
\ce{H + NO -> OH + N}
\label{eq:NO}
\end{equation}

\begin{figure}[htbp]
\centering
\includegraphics[width=0.9\linewidth, angle=0]{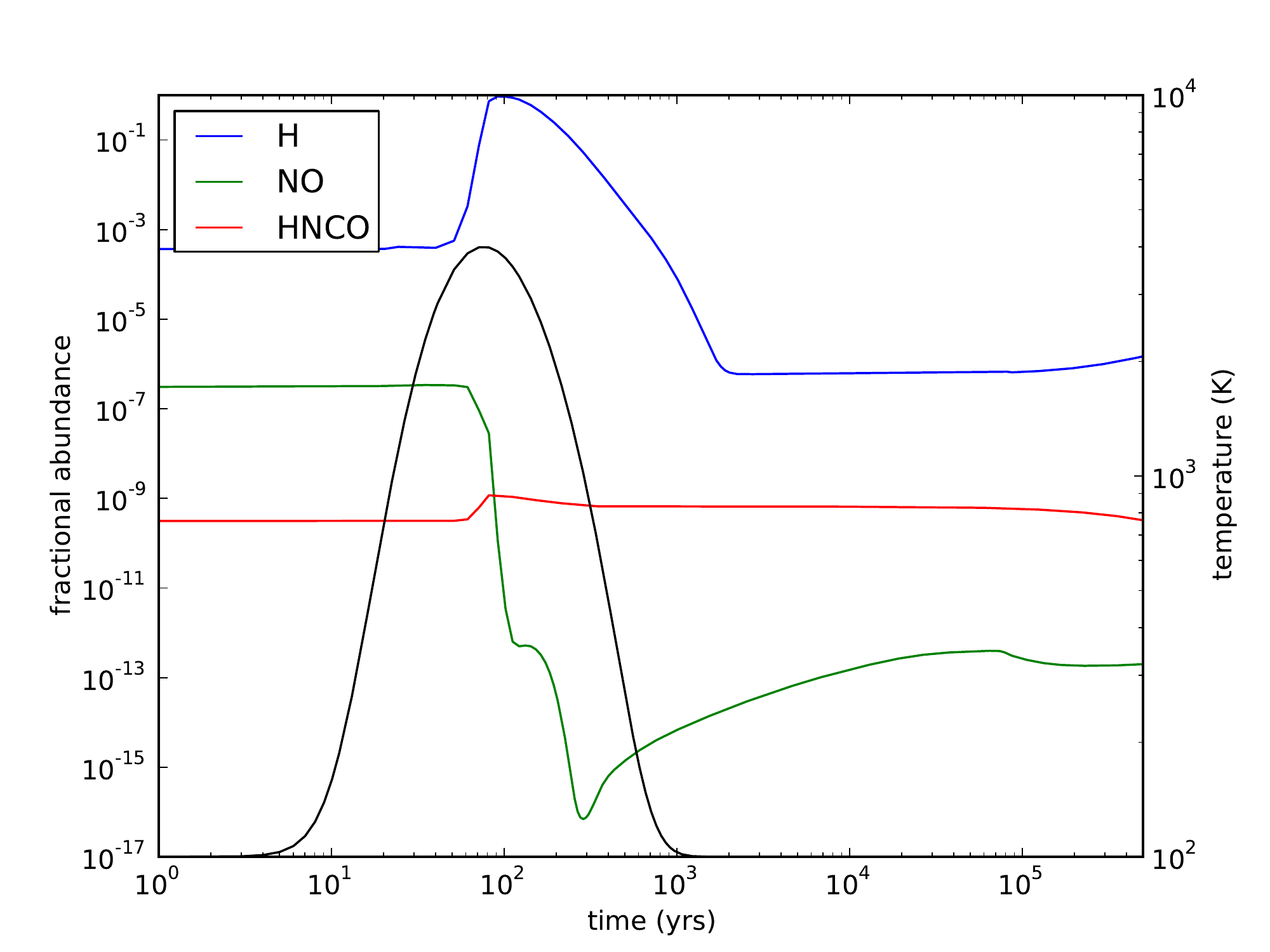}
\caption{Chemical shock modelling showing H, NO, and HNCO. This is fast (60 km/s) shock model 7. The rapid decrease in NO is due to reaction with H. HNCO requires NO to form and therefore does not increase in abundance after the shock, in contrast to what is found in both slow (20 km/s) and non shocked models.} 
\label{fig:HNCOform}
\end{figure}

This can be seen clearly in Figure ~\ref{fig:HNCOform}. Reaction ~\ref{eq:NO} only proceeds at very high temperatures and does not occur at all during the slow shock or without a shock. This is confirmed by the fact that in a slow shock model HNCO does in fact increase at late times as much as in a non shocked model.  SiO on the other hand clearly needs a fast shock to be enhanced and the level of enhancement will depend on the initial elemental depletion. Clearly there are too many parameters that play a role for us to be able to quantitatively fit the HNCO and SiO observed in the four different locations; nevertheless we do have an explanation for the anti-correlation of these two species: a high SiO and a low HNCO seem to indicate the presence of a fast shock while a low SiO and a high HNCO imply either a very slow shock or a warm dense non shocked gas.  

\begin{table}[htbp]
\begin{center}
\caption{Chemical shock model input parameters}
\label{tab:s-models}
\begin{tabular}{c c c c c c}
Model &  n(H$_{2}$)  & V$_{s}$ & T$_{max}$  & t$_{sat}$ & Si  \\
number & (cm$^{-3}$) & (km s$^{-1}$)  & (K) & (yr) & Depletion\\
\hline
1 & \num{e3} & 60 & 4000 & 380 & 1\% Solar\\
2 & \num{e3} & 60 & 4000 & 380 & 0.1\% Solar \\
3 & \num{e3} & 60 & 4000 & 380 & 0.01\% Solar \\
4 & \num{e4} & 60 & 4000 & 38 & 1\% Solar \\
5 & \num{e4} & 60 & 4000 & 38 & 0.1\% Solar \\
6 & \num{e4} & 60 & 4000 & 38 & 0.01\% Solar \\
7 & \num{e5} & 60 & 4000 & 3.8 & 1\% Solar \\
8 & \num{e5} & 60 & 4000 & 3.8 & 0.1\% Solar \\
9 & \num{e5} & 60 & 4000 & 3.8 & 0.01\% Solar \\
10 & \num{e3} & 20 & 900 & 570 & 1\% Solar\\
11 & \num{e3} & 20 & 900 & 570 & 0.1\% Solar \\
12 & \num{e3} & 20 & 900 & 570 & 0.01\% Solar \\
13 & \num{e4} & 20 & 900 & 57 & 1\% Solar \\
14 & \num{e4} & 20 & 900 & 57 & 0.1\% Solar \\
15 & \num{e4} & 20 & 900 & 57 & 0.01\% Solar \\
16 & \num{e5} & 20 & 900 & 5.7 & 1\% Solar \\
17 & \num{e5} & 20 & 900 & 5.7 & 0.1\% Solar \\
18 & \num{e5} & 20 & 900 & 5.7 & 0.01\% Solar \\

\end{tabular}
\end{center}
\end{table}

\begin{table}[htbp]
\begin{center}
\caption{Chemical non-shock model input parameters}
\label{tab:ns-models}
\begin{tabular}{c c c}
Model &  n(H$_{2}$)  & Si  \\
number & (cm$^{-3}$)  & Depletion\\
\hline
19 & \num{e4}  & 1\% Solar \\
20 & \num{e4} &  0.1\% Solar \\
21 & \num{e4} & 0.01\% Solar \\
22 & \num{e5}  & 1\% Solar \\
23 & \num{e5} &  0.1\% Solar \\
24 & \num{e5} & 0.01\% Solar \\
\end{tabular}
\end{center}
\end{table}

\section{Conclusions}

We have used the Plateau de Bure Interferometer to map two shock tracers, SiO and HNCO. SiO$(3-2)$ is detected strongly to the East of the AGN and to some extent to the West. HNCO$(6-5)$ is detected more strongly to the West, but is also detected in the East. The emission of the two lines is slightly offset from one another. We extracted spectra to analyse from the four peak emission locations of both lines. This allowed us to complete a RADEX radiative transfer modelling using our observations and SiO$(2-1)$ and HNCO$(5-4)$ from literature. We used parameters for the East and West Knot obtained by \citet{2014A&A...570A..28V} through a modelling of HCN, CS, CO and HCO$^{+}$. We found that in order to obtain a fit to our observations, the gas density, n(H$_2$) must be higher than \num{e4} cm$^{-3}$. We also found that, in general over the four locations, it was very hard to constrain a temperature. This may indicate that the gas as traced by HNCO and SiO is not at a constant temperature, consistent with a shocked region's varying temperature. Although this could also be due to the gas being of higher temperature than the upper energy levels of our transitions. 

In order to further investigate the origin of the SiO and HNCO emission we completed chemical modelling. We modelled a representative fast shock (60 km s$^{-1}$), slow shock (20 km s$^{-1}$) and no shock. We found that SiO is significantly enhanced during the fast shocks, due to grain core sputtering of Si. It was slightly enhanced during the slow shock and was also produced to some extent in the no shock models. We found that HNCO actually decreased in the fast shock models due to the destruction of its precursor, NO. This occurs through reaction with atomic hydrogen and only proceeds at the very high temperatures found during the fast shock. To confirm this, during the slow shock, HNCO abundance significantly increases.  HNCO also increased in abundance without need for a shock, in warm dense gas. This leads us to conclude that a high SiO but low HNCO abundance are indicative of a fast shock, whereas a low SiO and high HNCO abundance may indicate the presence of a slow shock, or of warm dense non shocked gas. Observations of the East Knot seem to therefore suggest gas in the region is heavily shocked. The offset of the HNCO peak to the SiO peak suggests that there may be regions in the East Knot away from the main shock that are undergoing a milder shock (particularly around our East Knot 2). The weak SiO emission and stronger HNCO emission in the West Knot suggests that there are not fast shocks occurring. There may be slower shocks, or the gas may be warm, dense and non-shocked. The results of our RADEX analysis, where we struggle to constrain temperature in the West Knot, point to the milder shocks as the more likely solution.
\FloatBarrier

\section{Acknowledgements}

This work was supported by the Science \& Technology Facilities Council (STFC). SGB acknowledges support from Spanish grants AYA2012-32295 and AYA2013-42227-P.  AF thanks the Spanish MINECO for funding support from grants CSD2009-00038, FIS2012-32096, AYA2012-32032, and ERC under ERC-2013-SyG, G. A. 610256 NANOCOSMOS.

\bibliographystyle{aa}
\bibliography{ngc1068_PdB_v9}

\end{document}